\documentclass[a4paper,11pt]{article}
\usepackage[utf8]{inputenc}
\usepackage[english]{babel}
\usepackage{braket}
\usepackage{xspace}
\usepackage{mathdots}
\usepackage{stackrel}
\usepackage[dvipsnames]{xcolor}
\usepackage{appendix}

\usepackage{mathtools}
\usepackage{bbm}
\usepackage{comment}
\usepackage{subfigure}
\usepackage[T1]{fontenc}               
\usepackage[left=3cm,
            right=2.5cm,
            top=2.5cm,
            bottom=3cm
            ]{geometry}                
\usepackage{graphicx}
\usepackage{mathrsfs}
\usepackage{appendix}
\usepackage{fancyhdr}
\usepackage{amsmath,                   
            amssymb,                   
            amsthm}                    
\DeclareMathOperator{\sech}{sech}
\usepackage{setspace}
\usepackage{cancel}

\usepackage{cellspace}
\usepackage[margin=30pt, bf, font=small, center, justification=justified]{caption}[2004/07/16]
\usepackage[numbers,sort&compress]{natbib}
\usepackage{doi}
\usepackage[
]{hyperref}

\definecolor{softred}{rgb}{0.7,0.3,0.2} 
\definecolor{softorange}{rgb}{0.8,0.4,0.1}

\title{\bf Long-range deformations in Gaussian states}

\date{}

\begin{document} 

\maketitle
\begin{center} Francisco Pereira$^{1}$, Nandagopal Manoj$^{2}$, Sara Murciano$^{1}$
\end{center}
\medskip 
{\small
  $^{1}$ 
  Universit\'e Paris-Saclay, CNRS, LPTMS, 91405, Orsay, France.  }\\
{\small{$^{2}$} Department of Physics and IQIM, Caltech,
    1200 E. California Bl., Pasadena, CA 91125, USA.} 

\begin{abstract}
Imaginary-time evolution by a local Hamiltonian cannot induce a phase transition in one dimension, but longer-range interactions may subvert such constraints. Starting from the ground state of the Kitaev Majorana chain, we modify the wave function by an imaginary-time evolution generated by a quadratic Hamiltonian with power-law couplings that enhance pairing correlations, typically of the form $1/r^{\alpha}$, where $r$ is the distance between two sites. As the state remains Gaussian, entanglement and correlation functions can be computed analytically. We find that the decay exponent $\alpha$ controls three distinct infrared regimes: for $\alpha>1$, the deformation produces only smooth crossovers at finite deformation strength, while the topological regime is reached only asymptotically as the deformation strength tends to infinity. At $\alpha=1$, the deformation induces an immediate flow to the topological phase: an infinitesimal deformation strength drives the system to a topological regime, and in a particular case, an emergent Kramers-Wannier symmetry enforces Ising-like scaling at long distances. For $\alpha<1$, the deformed state shows the same critical-like behavior for all non-zero deformation strength. We observe that even with arbitrarily long-range interactions, these models do not display a sharp phase transition at non-zero deformation strength.

\end{abstract}

 \tableofcontents

 \newpage

\section{Introduction}

In quantum many-body systems, it is often insightful to start from a simple reference state and deform the wave function in a local basis to construct families of wave functions that continuously interpolate between different regimes, while often preserving enough structure to remain analytically tractable. These constructions can realize novel types of phase transitions, such as conformal quantum critical points~\cite{Ardonne2004}.
From this viewpoint, the deformation can be understood as a reweighting of the wave function coefficients according to a local functional of the associated configuration~\cite{Castelnovo_from_2005, Castelnovo2008}. 
Such a construction is especially powerful when the amplitudes are real and nonnegative in a local basis, as they can often be associated with the Boltzmann weight of a classical configuration and we can import powerful techniques from statistical physics~\cite{Rokhsar_superconductivity_1988, Henley_from_2004}. As a result, the normalization of the quantum state and a broad class of equal-time observables can be mapped onto the partition function and correlation functions of an associated classical statistical-mechanics model. 
This quantum-to-classical dictionary provides direct access to the universal features encoded in the wave function and often makes it possible to identify critical points and characterize the competing forms of order in a transparent way. A further motivation for considering such deformations, which is also the main idea of the present work, is that the deformation parameter naturally acts as a continuous tuning knob potentially driving the system from one phase to another. For example, it may interpolate between a topologically ordered phase and a conventionally ordered, non-topological phase~\cite{Castelnovo2008,Castelnovo2010}. Such models offer a direct wave function-based perspective on the corresponding phase transition which, although fine-tuned as a traditional quantum critical point~\cite{IsakovFendley2011}, lies beyond the Landau paradigm.

In this work, we use the ground state of the Kitaev Majorana chain~\cite{Kitaev_unpaired_2001, Alicea_new_2012} as the reference wave function, whose free-fermion (Gaussian) nature makes its correlation functions and entanglement structure analytically tractable. We then introduce a long-range deformation, implemented as a reweighting of the wave function amplitudes via imaginary time-evolution by a free-fermion Hamiltonian with power-law decaying fermion bilinear terms, with decay exponent $\alpha$~(see the reviews \cite{Reviewstatic,DEFENU20241} and references therein). The deformation acts as a controlled modification of the Gaussian state, and its strength provides a natural tuning parameter to explore the evolution from an exactly solvable Gaussian limit to regimes with enhanced correlations, long-range order, or critical-like behavior. 
In particular, we ask whether starting from a trivial Gaussian state, such a Gaussian, long-range deformation can trigger a transition to a topologically non-trivial phase. 
We find that the choice of power-law exponent $\alpha$ dictates the universal long-distance features of the deformed wave function (see Fig.~\ref{fig:phasediagram}), similar to other studies of long-range interacting systems, including ground states of long-range interacting Hamiltonians~\cite{vodola2014,vodola_long-range_2015,Gong_topological_2016} and steady states of quantum circuit dynamics~\cite{Minato_fate_2022,Block_measurement-induced_2022,Muller_measurement-induced_2022}.
We also study the stability of the critical Gaussian state to such deformations.

\subsection*{Summary of the main results}
More concretely, we start from a Gaussian state $\ket{\psi}$, critical or otherwise, and we deform it via the non-unitary operator $e^{\beta H}$, where $H$ is a translation-invariant long-range Hamiltonian quadratic in fermion operators. The couplings entering the non-unitary deformation connect degrees of freedom at arbitrarily large distances, whose strength decays with the separation $r$ as a power law, typically of the form $1/r^\alpha$. The exponent $\alpha$ is therefore the parameter that controls how extended the deformation is in space: large $\alpha$ means that distant interactions are strongly suppressed, so the deformation is effectively close to short-ranged, while small $\alpha$ means that far-apart sites remain significantly coupled, so the deformation is genuinely long-ranged. In physical terms, decreasing $\alpha$ makes the generator of imaginary-time evolution increasingly nonlocal and this has a direct impact on the long-distance properties of the many-body state. With this intuition in mind, we identify three qualitatively different regimes. 

For $\alpha>1$, the long-range tail is relatively weak: the deformation still acts over all distances, but the couplings decay fast enough that the many-body state is not drastically reorganized at finite deformation strength $\beta$. If the starting point is already a critical state, the deformation gradually weakens that critical-like behavior in terms of entanglement or correlations. If, instead, the initial state is noncritical, the deformation can drive it toward a topological regime, but only in a smooth way. The system therefore does not undergo a sharp transition at finite $\beta$; rather, it approaches the new regime asymptotically. Physically, this means that when $\alpha>1$, the long-range character of the deformation is not strong enough to overcome the original short-range structure of the state at any finite $\beta$.

The point $\alpha=1$ marks the threshold between weak and strong long-range behavior. Here, the tail of the interaction decays just slowly enough to compete seriously with the local structure of the original system. As a result, it can happen that the system develops scale-invariant behavior such as logarithmic growth of entanglement entropy and power-law correlation functions over a broad range of parameters, even if the initial state is in the trivial gapped phase.

In other cases, the same decay can produce an immediate flow into a topological phase: as soon as $\beta>0$, the long-distance properties are governed by a topological infrared behavior.
Moreover, when deforming a trivial gapped state, at a certain value of $\beta$ the system develops an emergent Kramers-Wannier symmetry which we make precise within the framework of free-fermionic Gaussian states. We show that this emergent symmetry strongly constrains the long-distance properties of the state to match that of the Ising conformal field theory.

For $\alpha<1$, the long-range character of the deformation becomes dominant. Since the decay with distance is now so slow, even very distant parts of the system remain significantly coupled by the deformation. This has a dramatic consequence: as soon as the deformation is switched on, even for arbitrarily small $\beta$, the state develops the hallmarks of critical behavior at large distances. For example, the decay exponent of power-law correlations immediately saturates to a $\beta$-independent value. In this regime, the non-unitary evolution is no longer continuously modifying a pre-existing phase, but is directly generating a new long-distance universality class. The key message is therefore that sufficiently slowly decaying long-range couplings can generate signatures of criticality without requiring the system to be tuned to a conventional equilibrium critical point.

\begin{figure}[t!]
\centering
\includegraphics[width=.9\textwidth]{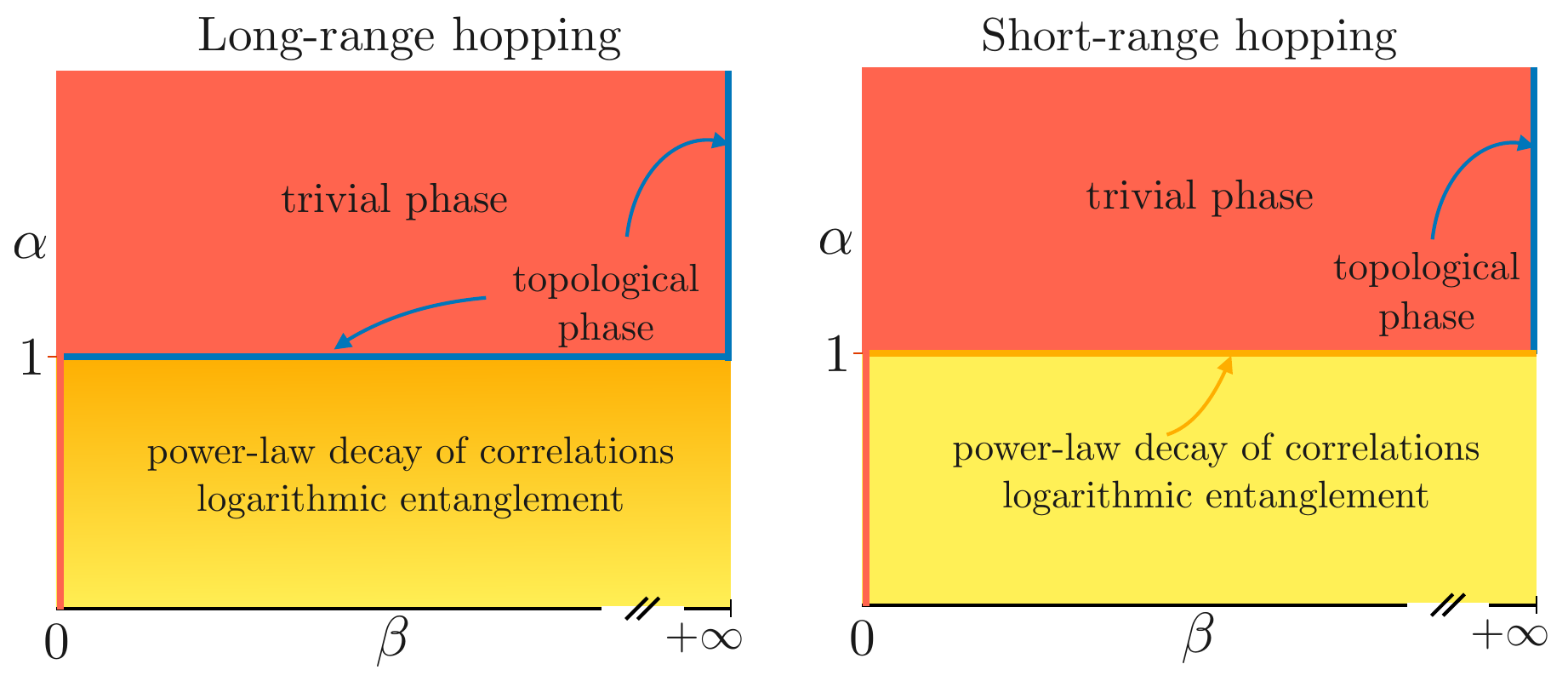}
\caption{\textbf{Phase diagram of a deformation of a trivial state.} We plot the phase diagram as a function of the measurement strength, $\beta$, and power-law decaying pairing terms (decay exponent $\alpha$), for both long-range and short-range hopping deformations (see Eq.~\eqref{eq:Hev}). Starting from the trivial phase, we distinguish whether the system is in a trivial or topological phase as we turn on the deformation, or if the system exhibits power-law correlations or logarithmic scaling of the entanglement.}
\label{fig:phasediagram}
\end{figure}

The overall picture is thus quite intuitive. The exponent $\alpha$ measures how efficiently the deformation can transmit correlations across long distances. When $\alpha$ is large, the deformation remains close to local and its effect is smooth. When $\alpha=1$, the system sits at the boundary between local and genuinely nonlocal behavior, displaying continuously varying long-distance behavior. When $\alpha<1$, the deformation becomes so extended that it can immediately reshape the infrared physics of the state. From this perspective, the paper shows that the emergence of critical-like behavior is controlled not simply by the strength $\beta$ of the deformation, but more fundamentally by how slowly its spatial tail decays, namely by the value of $\alpha$. We summarize these results in Fig.~\ref{fig:phasediagram}. We have in mind an initial state in a trivial phase (i.e. exponential decay correlations and area law entanglement) and we consider Hamiltonian with long-range pairing, but long-range or short-range hopping.

We proceed in Sec.~\ref{sec:review} by first reviewing the model and detailing our protocol. Once we specify the Gaussian reference state and the long-range deformation preserving the Gaussianity of the state, we analytically compute the entanglement and the correlations of the deformed wave function in Secs.~\ref{sec:entanglement} and~\ref{sec:correlations}, respectively. We further support our results through the entanglement spectrum in Sec.~\ref{sec:spectrum}. We collect all the analytical details in the appendices.

\section{Long-range non-Hermitian evolution}\label{sec:review}
The main model we consider in this work is the following. We start from the ground state of the anisotropic Kitaev chain  of length $N$, which in fermionic
variables reads
\begin{equation}\label{eq:init}
    H_{0} = - \frac{1}{2}\sum_{j=1}^{N-1} \left( c_j^\dagger c_{j+1}  + \gamma \, c_j^\dagger c_{j+1}^\dagger + {\rm h.c.} \right)
+ h \sum_{j=1}^N c_j^\dagger c_j 
- \frac{1}{2} \left( c_N^\dagger c_1 + \gamma \, c_N^\dagger c_1^
\dagger+ {\rm h.c.}  \right),
\end{equation}
where $c_j^{\dagger},c_j$ are fermionic creation and annihilation operators acting on the site $j$, satisfying the canonical anticommutation relations $\{c_j,c_{j'}^{\dagger}\}=\delta_{jj'}$, $\{c_j^{\dagger},c_{j'}^{\dagger}\}=\{c_j,c_{j'}\}=0$.
Here, $\gamma$ is the anisotropy parameter, $h$ ($h>0$) is the transverse magnetic field, and we work in the odd-parity sector, i.e., we impose periodic boundary conditions $c_{1+N}=c_1$. The non-unitary evolution of this state is governed by a
quadratic, translational invariant Hamiltonian~(see Refs.~\cite{vodola2014,vodola_long-range_2015} for the original introduction of the long-range Kitaev chain, and~\cite{Alecce2017,ares2018ent,ares2015ent,Giuliano2018,Jager2020,Cai2017,Solfanelli2023,solfanelli_quantum_2023,Cinnirella2025,Baghran24,AgarwalJoglekar2021,Fraxanet2022,Francica2022,mishra_disordered_2020,hernandez-santana_correlation_2017,pezze_multipartite_2017,maghrebi_causality_2016,defenu_universal_2019,uhrich_out--equilibrium_2020,van_regemortel_information_2016,jaschke_critical_2017,lepori_singular_2017} for later extensions and variations) 
\begin{equation}\label{eq:Hev}
H = \frac{1}{2} \sum_{n=1}^{N} \sum_{|l|<N/2} 
\left( 2A_{l} c_{n}^{\dagger} c_{n+l} + B_{l} c_{n}^{\dagger} c_{n+l}^{\dagger} - B_{l} c_{n} c_{n+l} \right).
\end{equation}
We mainly focus on two possible choices for $A_l,B_l$:
\begin{itemize}
    \item The short-range hopping, long-range pairing model $H_{\text{SH}}$: where $A_l=-\frac{t}{2}\delta_{|l|,1}+\mu\delta_{l,0}$, where $\mu$ is the 
    chemical-potential strength, and $B_l=\frac{l}{2|l|^{\alpha+1}}$;
    \item The long-range hopping, long-range pairing model $H_{\text{LH}}$: where $A_l=\frac{1}{2|l|^{\alpha}}+\mu\delta_{l,0}$ and 
    $B_l=\frac{l}{2|l|^{\alpha+1}}$.
\end{itemize}

We now rewrite the Hamiltonian~\eqref{eq:Hev} in a diagonal basis. 
We first introduce the Fourier modes $c_k=\frac{e^{-i\pi/4}}{\sqrt{N}}\sum_{n=1}^N e^{-ik n    }c_n$,

with $ k \in \Omega=\{0,\frac{2\pi}{N},\cdots,\frac{2\pi}{N}(N-1)\}$, so that, up to an overall constant, the Hamiltonian can be rewritten as (we consider even $N$ and we neglect an additive constant) 
\begin{equation}\label{eq:HevF}
H 
=\sum_{k\in \Omega'} \begin{pmatrix}
    c_k^{\dagger}& c_{2\pi-k}
\end{pmatrix}
\begin{pmatrix}
    G_k& F_k\\ F_k&-G_k
\end{pmatrix}
\begin{pmatrix}
    c_k\\ c_{2\pi -k}^{\dagger}
\end{pmatrix}
,\quad G_k=\sum_l A_le^{ikl},\quad F_k= (-i)\sum_l B_le^{ikl},
\end{equation}
where $\Omega'=\{0,\frac{2\pi}{N},\cdots,\pi\}$.
By doing a Bogoliubov transformation
\begin{equation}
    \begin{pmatrix}
    \psi_k\\\psi_{2\pi-k}^{\dagger}
\end{pmatrix}
=\begin{pmatrix}
    \cos\eta_k&\sin\eta_k\\
    -\sin\eta_k&\cos\eta_k
\end{pmatrix}\begin{pmatrix}
    c_k\\c_{2\pi-k}^{\dagger}
\end{pmatrix},\label{bogo}
\end{equation}
where 
\begin{equation}
    \cos (2\eta_k)=\frac{G_k}{\sqrt{G_k^2+F_k^2}},\quad\quad\quad \sin(2\eta_k)=\frac{F_k}{\sqrt{G_k^2+F_k^2}}.
\end{equation}
We can diagonalize the Hamiltonian as
\begin{equation}   H=\sum_{k=0}^{N/2}\epsilon_k(
    \psi_k^{\dagger} \psi_{k}
   -\psi_{2\pi-k} \psi_{2\pi-k}^{\dagger}) \label{diagH}
\end{equation}
with dispersion relation $\epsilon_k=\sqrt{G_k^2+F_k^2}$. In the thermodynamic limit, $N\to \infty$, the energy density 
\begin{equation}
    e=\int_{0}^{2\pi} \frac{dk}{2\pi} \epsilon_k 
\end{equation}
remains finite for all $\alpha>0$ (also for $\alpha<1$, since $\epsilon_k \sim |{k}|^{\alpha-1}$ for small $k$). Therefore, we do not need a system-size-dependent normalization factor in the Hamiltonian~\eqref{eq:HevF} 
to have a well-defined thermodynamic limit, even when $\alpha < 1$. 
Notice that this is a feature of Gaussian states, and it is not true for long-range interacting systems, where the energy density is not finite anymore. 

Having in mind the structure of the Hamiltonian of the initial state \eqref{eq:init} and of its evolution \eqref{eq:Hev}, we are now ready to define the details of our protocol. We are interested in the long-range deformation 
\begin{equation}\label{eq:longrange}
    \ket{\psi(\beta)}=\frac{1}{\sqrt{\mathcal{N}}}e^{\beta H}\ket{\psi(0)},
\end{equation}
where $\ket{\psi(0)}$ is the ground state of Eq. \eqref{eq:init} and $\mathcal{N} = \bra{\psi(0)}e^{2\beta H}\ket{\psi(0)}$ is an overall normalization constant due to the non-Hermiticity of the deformation. We choose the deformation strength to be $\beta>0$. By Fourier transforming the fermionic variables,
we can rewrite the Hamiltonian \eqref{eq:init} as 
\begin{equation}\label{eq:HinitF}
H_{0}=\sum_{k=0}^{N/2} \left[
    (\cos (2\pi k/N)-h)\left( c_{2\pi-k}c^\dagger_{2\pi-k} 
     - c^\dagger_kc_k
     \right)
     -\gamma\sin (2\pi k/N)
     \left( 
     c_{2\pi-k}c_{k}
    +c^\dagger_kc^\dagger_{2\pi-k}  
    \right)
    \right].
\end{equation}

The first step in solving the problem is to determine the ground state of Eq.~\eqref{eq:init}. 
Since $H_{0}$ is a sum over the positive momenta and we have imposed periodic boundary conditions (i.e., the ground state lives in the odd parity sector), the ground state decomposes in momentum space as
\begin{equation}
\label{eq:factorized_GS}
|\psi(0)\rangle=c_0^\dagger\ket{0} \otimes \bigotimes_{k=1}^{N/2}|\psi_k(0)\rangle,\quad 
|\psi_k(0)\rangle\equiv (u_k(0) + v_k(0)c_{k }^{\dagger} c_{2\pi-k }^{\dagger} ) |0\rangle.
\end{equation}
The tensor product notation is unambiguous as each state $\ket{\psi_k}$ is in a definite fermion parity sector. The coefficients are given by
\begin{equation}\label{eq:initial}
    u_k(0)=\frac{1}{\sqrt{2}}\sqrt{1+\frac{h-\cos (k)}{\omega_k}}, \quad  
   v_k(0)=\frac{\gamma\sin (k)}{\sqrt{2} \sqrt{\omega_k^2+(h-\cos (k))\omega_k}},
\end{equation}
and $\omega_k=\sqrt{(h-\cos (k))^2+\gamma^2\sin
   ^2(k)}$. Since the Hamiltonian \eqref{eq:HevF} retains the same structure as Eq.~\eqref{eq:HinitF} in the momentum space, the non-unitarily evolved state maintains the factorized form
\begin{equation}
\label{eq:timeevolved_state}
| \psi(\beta)\rangle=c^{\dagger}_0\bigotimes_{k} \vert \psi_k(\beta)\rangle=
c^{\dagger}_0\bigotimes_{k}
\left(\frac{u_k(\beta)+v_k(\beta)c^\dagger_{k} c^\dagger_{2\pi-k}}{\sqrt{| u_k(\beta)\vert^2+\vert v_k(\beta)\vert^2}}\right)| 0\rangle 
\,,
\end{equation}
where the denominator ensures the proper normalization of the state. To determine the post-measurement coefficients $u_k(\beta)$ and $v_k(\beta)$, we need to solve the Schrödinger equation 
\begin{align}
    \frac{d}{d\beta}\ket{\psi_k(0)}=
    H_k\ket{\psi_k(0)}\implies
    \frac{d}{d\beta} \begin{pmatrix}
       u_k(\beta)\\
        v_k(\beta)
    \end{pmatrix}=
    \begin{pmatrix}
            -G_k& F_k\\
            F_k& G_k
    \end{pmatrix}
    \begin{pmatrix}
        u_k(\beta)\\
        v_k(\beta)
    \end{pmatrix}.
\end{align}
By solving the differential equation with the initial condition fixed by Eq. \eqref{eq:initial}, we get 
\begin{equation}
    \begin{split}
        u_k(\beta)=&\,\left(\cosh(\beta\epsilon_k)u_k(0) -\frac{\sinh(\beta\epsilon_k) (\beta  \epsilon_k)  
   (G_k u_k(0)-F_k v_k(0))}{\epsilon_k}\right),\\
   v_k(\beta)=\, & \left(\cosh(\beta\epsilon_k)v_k(0) +\frac{\sinh (\beta  \epsilon_k)
   (G_k v_k(0)+F_k u_k(0))}{\epsilon_k}\right),
    \end{split}
    \label{evolutionformula}
\end{equation}

We now have all the ingredients to describe the post-measurement state $|\psi(\beta)\rangle$. The main quantities we are interested in are the entanglement entropy and the correlation functions. 
Since the state is Gaussian, everything boils down to computing the two-point correlation matrix, expressed in terms of the fermionic operators $(c_j^\dagger, c_j)$ as 
\begin{equation}
\Gamma_{jj'}=
 \begin{pmatrix}
      2 \langle c_j
 c^\dagger_{j'}\rangle- \delta_{jj'}  && 2\langle c_j
 c_{j'}\rangle
        \\
    2 \langle c^\dagger_j
 c^\dagger_{j'}\rangle && 2 \langle c^\dagger_j
 c_{j'}\rangle- \delta_{jj'}
    \end{pmatrix}.
\end{equation}
We can use the explicit form of the state \eqref{eq:timeevolved_state} to compute the matrix elements, and in the thermodynamic limit $N\to \infty$  we get
\begin{equation}\label{eq:corr}
\Gamma_{jj'}= \int_{0}^{2\pi}\frac{dk}{2\pi}e^{ik(j-j')}\mathcal{G}^{\mathrm{Block}}(k,\beta),\qquad \mathcal{G}^{\mathrm{Block}}(k,\beta)=\begin{pmatrix}
    n(k,\beta)&g(k,\beta)\\
    g^*(k,\beta)&-n(k,\beta)
\end{pmatrix},   \end{equation}
where
\begin{equation}\label{eq:ng}
n(k,\beta) = \frac{|u_k(\beta)|^2 - |v_k(\beta)|^2}{|u_k(\beta)|^2 + |v_k(\beta)|^2}, \quad
g(k,\beta) = \frac{-2i u_k^*(\beta) v_k(\beta)}{|u_k(\beta)|^2 + |v_k(\beta)|^2}.
\end{equation}
Once we have determined the coefficients $u_k(\beta)$ and $v_k(\beta)$, we can plug them in Eq. \eqref{eq:ng} and we get the final expression for the correlators. In the next sections, we show how to apply standard techniques that relate the entanglement and nontrivial correlation functions to the correlation matrix~(\ref{eq:corr}). We use the same correlation matrix~(\ref{eq:corr}) also to compute the entanglement entropy and correlation functions numerically, working directly in the thermodynamic limit.

\section{Entanglement of the deformed wave function}\label{sec:entanglement}
In this section, we analyze the entanglement properties of our system using the Toeplitz-matrix formalism. We first briefly review the technical ingredients needed to compute the entanglement entropy from the symbol of a Toeplitz matrix. We then adapt this framework to the specific structure of the state~\eqref{eq:timeevolved_state}. Finally, we use this representation to study the behavior of the entanglement entropy in different parameter regimes, highlighting the qualitative differences between trivial and topological phases.
\subsection{Asymptotic entanglement entropy from Toeplitz symbols}

The correlation matrix \eqref{eq:corr} has a  Toeplitz structure, meaning that its entries depend only on the distance between sites, so every diagonal is constant. In Fourier space, this structure is encoded by the symbol $\mathcal{G}^{\mathrm{Block}}(k,\beta)$, a 
$2\times2$ matrix whose inverse Fourier transform yields Eq.~\eqref{eq:corr}. That is, the symbol $\mathcal{G}^{\mathrm{Block}}(k,\beta)$ is the equal-time momentum-space Green's function describing the Gaussian state. A Toeplitz matrix associated with a general $2\times 2 $ symbol $J_k$ is a $2\ell\times 2\ell$ block matrix with entries 

\begin{equation}
    [T_{\ell}(J)]_{jl}=\int_{0}^{2\pi}\frac{dk}{2\pi}e^{ik(j-l)}J_k, \quad j,l=1,\dots,\ell.
\end{equation}
This mathematical structure allows us to use powerful results on (block) Toeplitz matrices to extract asymptotic properties of the correlation matrix, like its determinant. For instance, a key result is the Fisher-Hartwig conjecture, according to which if $J_k$ is a $2\times 2$ symbol (as Eq. \eqref{eq:corr}) whose determinant has zero winding number around the origin and it has a finite set of discontinuities at momenta $k_1,...,k_r$, then the determinant of the $2\ell\times 2\ell$ block-Toeplitz matrix $T_{\ell}(J)$ 
behaves as
\begin{equation}\label{eq:conjecture1}
\log[\mathrm{det}\,T_{\ell}(J)]= \frac{\ell}{2\pi}\int_{0}^{2\pi} \log [\det J_k]dk +\frac{\log\ell}{4\pi^2}\sum_{\sigma=1}^r \mathrm{Tr}[\log J_\sigma^- (J_\sigma^+)^{-1}]^2+O(1).
\end{equation}
Here, $J^{\pm}_{\sigma}$ denotes the limit from the left and the right of the discontinuity $k_{\sigma}$. In the absence of discontinuities, the conjecture reduces to Szegő’s theorem~\cite{Its2005,DeiftItsKrasovsky2013}. The subleading $\log \ell$ contribution is the hallmark of the discontinuities of the symbol and is absent for a smooth symbol. This expression is particularly useful, as it enables us to access the asymptotic behavior of the entanglement. We derive a compact expression for the subsystem entanglement entropy by demonstrating that the symbol $\mathcal{G}^{\mathrm{Block}}(k,\beta)$ possesses a single discontinuity at $k^\ast = 0,\pi$; readers wishing to skip the derivation may proceed directly to Eq.~\eqref{genEntropy}.

Given the density matrix of the full system, $\rho = \ket{\psi(\beta)}\bra{\psi(\beta)}$, we obtain the reduced density matrix of the subsystem $A$ as $\rho_A = \mathrm{Tr}_B \rho$.
The R\'enyi entanglement entropy is then defined by
\begin{equation}
    S^{(n)}_A(\beta)=\frac{1}{1-n}\log\mathrm{Tr}\rho_A^n,\quad S^{(1)}_A(\beta)=\lim_{n\to 1}  S^{(n)}_A(\beta),
\end{equation}
where the replica limit $n\to 1 $ is the von Neumann entanglement entropy. For Gaussian states, the entanglement can be expressed only as a function of the correlation matrix~\cite{Peschel2009,Peschel2003}
\begin{equation}\label{eq:numericsent}
   S^{(n)}_A(\beta)=\frac{1}{1-n}\sum_{i=1}^{\ell} \log\left[ \left(\frac{1+\lambda_i}{2}\right)^n+\left(\frac{1-\lambda_i}{2}\right)^n \right],
\end{equation}
where $\lambda_i$ are the eigenvalues of $\Gamma$ restricted to the subsystem $A$ of size $\ell$. For the case where $A$ is a set of $\ell$ contiguous sites in a 1D translation invariant Gaussian state, we can apply Cauchy’s residue theorem to express $S^{(n)}_A$ as the complex integral 
\begin{equation}
   S^{(n)}_A(\beta)=  \frac{1}{4\pi i} \lim_{\varepsilon \to 1^+} \oint_{\mathcal{C}_\varepsilon} f_n(\lambda / \varepsilon) \frac{d}{d\lambda} \log \left[\det(T_{\ell}(J_k)) \right]\, d\lambda,
\label{complexint}
\end{equation}
where $J_k = \lambda I-\mathcal{G}(k,\beta)$, $\mathcal{C}_\varepsilon$ is a contour that surrounds the eigenvalues $\{\lambda_1, \dots, \lambda_\ell\}$, and
\begin{align}
f_n(\lambda) = \frac{1}{1-n}\log\left[\left( \frac{1 + \lambda}{2} \right)^n+ \left( \frac{1 - \lambda}{2} \right)^n\right].
\label{fentropy}
\end{align}
The integrand has singularities on the real axis within $[-1,1]$ and branch cuts extending along $(-\infty,-\varepsilon]\cup[\varepsilon,\infty)$. Therefore, we can study the symbol, and use the conjecture~\eqref{eq:conjecture1} to compute the entanglement entropy in the replica limit $n\to 1$.

We begin by noting that the factorized structure of the evolved state~\eqref{eq:timeevolved_state} is preserved under the dynamics, which implies $u_{2\pi-k}(\beta) = u_k(\beta)$ and $v_{2\pi-k}(\beta) = -v_k(\beta)$ by construction. It follows immediately that 
\begin{equation}\label{ngper}
    n(k,\beta)=n(2\pi-k,\beta),\quad \quad g(2\pi-k,\beta)=-g(k,\beta).
\end{equation}

\begin{figure}[h!]
\centering
\includegraphics[width=0.6\textwidth]{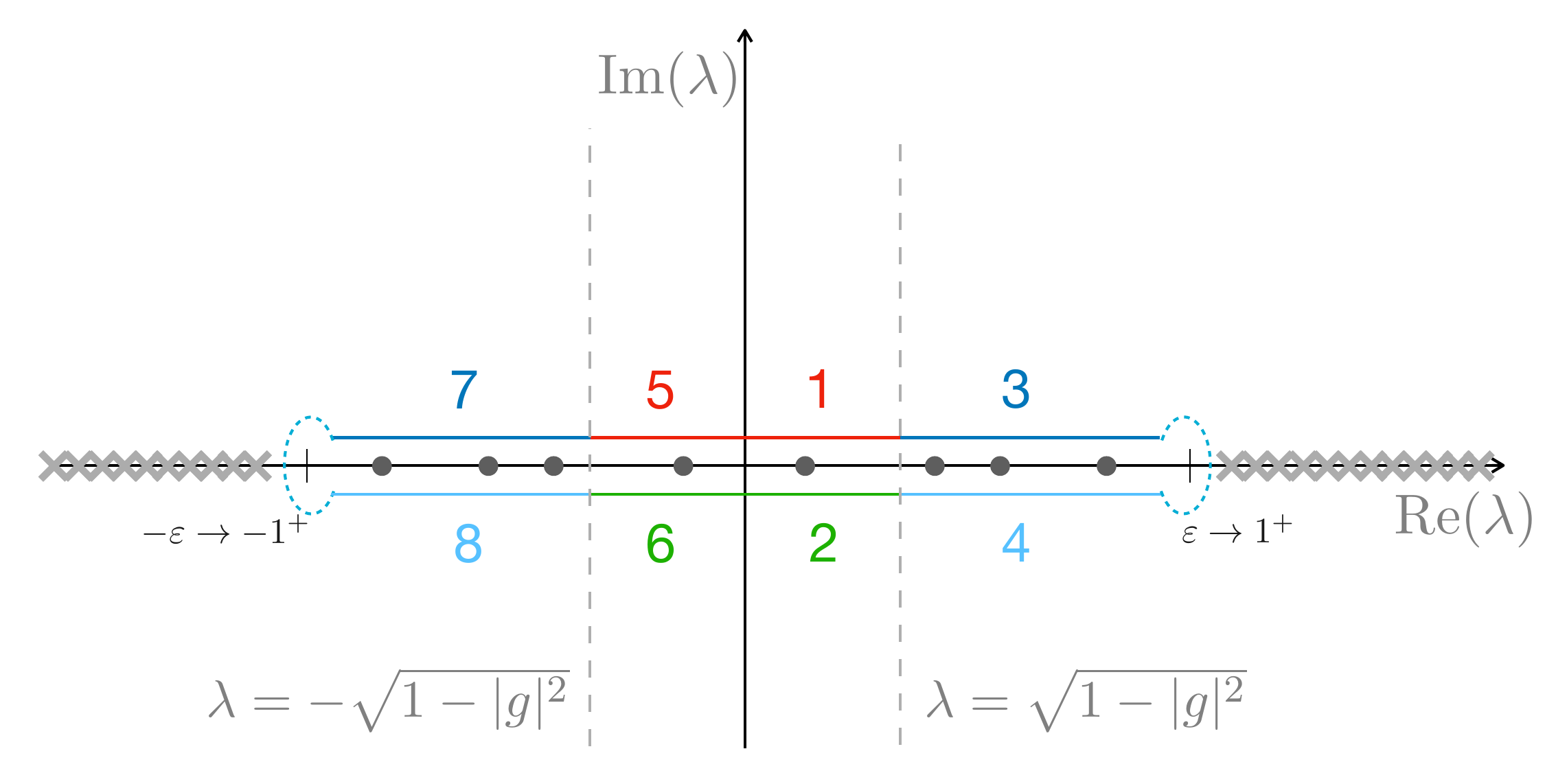}
\caption{Contour plot showing the eight sectors used in the analytic continuation of $\eta_\pm$. Regions $1$--$4$ correspond to $\lambda>0$, while regions $5$--$8$ correspond to $\lambda<0$. The two tables below list, for each numbered sector, which branch of $\eta_\pm$ is selected depending on whether the contour is taken on the upper or lower side of the cut.}
\label{fig:contour}
\vspace{0.8em}

\begin{minipage}[t]{0.47\textwidth}
\centering
\captionsetup{type=table}
\begin{tabular}{|c|c|c|}
\hline
$\displaystyle 1 + \frac{|g|^2}{\lambda^2 - 1}$ & Side of contour & $\eta_\pm$ expression \\
\hline
$1:\quad > 0$ & Top    & $\left( a \mp i b \right)$ \\
$2:\quad > 0$ & Bottom & $\left( a \pm i b \right)$ \\
$3:\quad < 0$ & Top    & $-i\left( a \pm b \right)$ \\
$4:\quad < 0$ & Bottom & $i\left( a \pm b \right)$ \\
\hline
\end{tabular}
\captionof{table}{\footnotesize Branches of $\eta_\pm$ for the sectors $1$--$4$ of Fig.~\ref{fig:contour}, i.e. the part of the contour plot associated with $\lambda>0$. The table shows how the sign of $1+|g|^2/(\lambda^2-1)$ together with the choice of the upper or lower side of the contour determines the corresponding expression.}
\end{minipage}
\hfill
\begin{minipage}[t]{0.47\textwidth}
\centering
\captionsetup{type=table}
\begin{tabular}{|c|c|c|}
\hline
$\displaystyle 1 + \frac{|g|^2}{\lambda^2 - 1}$ & Side of contour & $\eta_\pm$ expression \\
\hline
$5:\quad > 0$ & Top    & $\left( a \pm i b \right)$ \\
$6:\quad > 0$ & Bottom & $\left( a \mp i b \right)$ \\
$7:\quad < 0$ & Top    & $i\left( a \mp b \right)$ \\
$8:\quad < 0$ & Bottom & $-i\left( a \mp b \right)$ \\
\hline
\end{tabular}
\captionof{table}{\footnotesize Branches of $\eta_\pm$ for the sectors $5$--$8$ of Fig.~\ref{fig:contour}, corresponding to the part of the contour plot with $\lambda<0$. As in the left table, the contour side and the sign of $1+|g|^2/(\lambda^2-1)$ fix the branch that must be chosen in each region.}
\end{minipage}
\end{figure}

We have verified that, for the specific choice of the state~\eqref{eq:longrange}, $u_k(\beta)$ is a continuous function of $k$ for $k\in [0,2\pi]$, while $v_k(\beta)$ can have a discontinuity for $k=0$ or $k=\pi$. This means that $u_k(\beta)$ and $|v_k(\beta)|$ are continuous in $k$, while, in the presence of a discontinuity, $v_{k=0^+}(\beta)=-v_{k=2\pi^-}(\beta)$ and $v_{k=\pi^-}(\beta)=-v_{k=\pi^+}(\beta)$. Hence, the elements of the symbol for $k\to k_*$ take the form
\begin{equation}\label{eq:discontinuities}
n(k_*^+,\beta)=n(k_*^-,\beta)\equiv n(\beta),\qquad 
g(k_*^+,\beta)=-g_{k_*^-,\beta}(\beta)\equiv g(\beta),
\end{equation}
and they obey $|n(k,\beta)|^2+|g(k,\beta)|^2=1$. For notational simplicity, we omit the explicit $\beta$-dependence. The determinant of $J$ reads $\det (J_{\lambda})=\lambda^2-n^2(k,\beta)-|g(k,\beta)|^2=\lambda^2-1 $. Plugging Eq.~\eqref{eq:conjecture1} into \eqref{complexint} we get 
\begin{equation}
S_A^{(n)}(\beta) =\ell\frac{1}{4\pi i}\lim_{\varepsilon \to 1^+} \oint_{\mathcal{C}_\varepsilon} f_n(\lambda / \varepsilon)\frac{d}{d\lambda}\log(\lambda^2-1)+O(\log\ell).
\end{equation}
Since $f_n(\pm1)=0$, the integral above vanishes. If the symbol presents some discontinuities, to evaluate the logarithmic contribution to the entanglement entropy, we need to compute $J^- = \lambda I-\mathcal{G}(k_*^-,\beta)$ and $J^+ = \lambda I-\mathcal{G}(k_*^+,\beta)$. Explicitly, we get
\begin{align}
    J^-(J^+)^{-1}&=\frac{1}{\lambda^2-1}\begin{pmatrix}
        \lambda^2-1+2|g|^2&2g(\lambda-n)\\
        2g^*(\lambda+n)&\lambda^2-1+2|g|^2
    \end{pmatrix}.
\end{align}
The eigenvalues of this matrix are
\begin{equation}
\gamma_{\pm}=\left(\sqrt{1+\frac{|g|^2}{\lambda^2-1}}\pm\frac{|g|}{\sqrt{\lambda^2-1}}\right)^2 \equiv \eta_\pm^2.
    \label{eigen}
\end{equation}
Plugging this result and the conjecture~\eqref{eq:conjecture1} in Eq.~\eqref{complexint}, we get in the replica limit $n\to 1$
\begin{align}
    S_A^{(1)} = -\log\ell\left(\frac{1}{8\pi^3i}\right)\oint_\mathcal{C}d\lambda\log\left(\frac{1-\lambda}{1+\lambda}\right)\sum_{p=\{+,-\}}\log^2\left(\eta_p\right).
\end{align}
where $p$ labels the two eigenvalues of the matrix (\ref{eigen}). To transform this contour integral into a real integral, we encircle the interval $[-1,1]$, thereby collecting real contributions from both the upper and lower edges of the branch cut.
We then divide the contour into eight parts, distinguished by the following three different conditions:
\begin{itemize}
    \item $ \lambda>0$ or $\lambda<0$ 
    \item $1+\frac{|g|^2}{\lambda^2-1}>0$ or $1+\frac{|g|^2}{\lambda^2-1}<0$
    \item the position along the contour: upper or lower branch.
\end{itemize}
We define
\begin{equation}\label{abDef}
a = \sqrt{\left|1+\frac{|g|^2}{\lambda^2-1}\right|}, \hspace{2 cm} b = \frac{|g|}{\sqrt{|\lambda^2-1|}},
\end{equation}
and Tables 1-2 summarize all eight possible combinations. See also Fig.~\ref{fig:contour} for their graphical representation. 

Since the contour is oriented counterclockwise, the contribution from the upper edge of the cut enters with a minus sign. As a result, the first two rows of each table cancel pairwise, leaving only the sectors where $1 + \frac{|g|^2}{\lambda^2 - 1} < 0 .$
Summing the non-vanishing contributions we get
{\small
\begin{align}S_A^{(1)}(\beta) &= \frac{c_{\mathrm{eff}}}{3}\log\ell+O(1), \quad c_{\mathrm{eff}}=\left(\frac{3}{2\pi^2}\right)\int_{\sqrt{1-|g|^2}}^1d\lambda\log\left(\frac{1-\lambda}{1+\lambda}\right)\log\left(\frac{b-a}{b+a}\right).
\label{genEntropy}
\end{align}}

The effective central charge, $c_{\rm eff}$, is the coefficient of the logarithmic term, and it characterizes how the entanglement changes continuously as one moves away from $\beta=0$. Here $|g|$ is half of the jump in the symbol across the discontinuity $k^\ast = 0$ as defined in Eq.~\eqref{eq:discontinuities} and $a,b$ are defined in Eq.~\eqref{abDef}.

Equation~\eqref{genEntropy} provides a compact analytic expression for the entanglement entropy, depending solely on the function $g(k,\beta)$.

\subsection{Long-range regimes of non-unitary entanglement growth}

Before discussing the different long-range regimes, let us recall that logarithmic entanglement growth can arise only when the symbol of the evolved state develops a discontinuity. As follows from Eq.~\eqref{evolutionformula}, the evolved symbol at finite $\beta$ is built from the initial symbol and from the functions $F_k$ and $G_k$ generated by the deformation. Therefore, at finite $\beta$ a 
discontinuity can appear only if it is already present in the initial 
state, or if $F_k$ or $G_k$ is itself discontinuous in $k$. Physically, this means that the non-unitary deformation cannot create a logarithmic term out of a completely smooth input unless the deformation itself introduces a sharp momentum-space structure. In particular, when the initial state is gapped and the deformation is smooth in momentum space, the symbol remains smooth for all finite $\beta$, and the entanglement entropy obeys an area law throughout the evolution. By contrast, if the deformation drives the state toward a `projection state' (the state at $\beta \to \infty$)

whose symbol has a finite jump, then this discontinuity is inherited by the evolved state already at any $\beta>0$, and the entropy acquires a logarithmic contribution whose coefficient is fixed by the jump through Eq.~\eqref{genEntropy}. The role of the evolution is then only to renormalize the magnitude of this discontinuity. Note that the $L \to \infty$ and $\beta \to \infty$ limits commute---the different momentum sectors decouple and the problem reduces to non-unitary deformations in a finite dimensional Hilbert space due to its translation invariant free fermion nature. 

For the sake of simplicity, we mainly focus on $|\mu|<  |t|$ for evolution under $H_{\rm SH}$ and $-\sum_l \frac{1}{l^{\alpha}}<\mu<  \sum_l \frac{(-1)^{l+1}}{l^{\alpha}}$ for deformation under $H_{\rm LH}$, and we refer to the Appendices~\ref{app:Hev} and~\ref{app:symbol} for more details about the remaining cases.

\begin{figure}[t!]
\includegraphics[width=.47\textwidth]{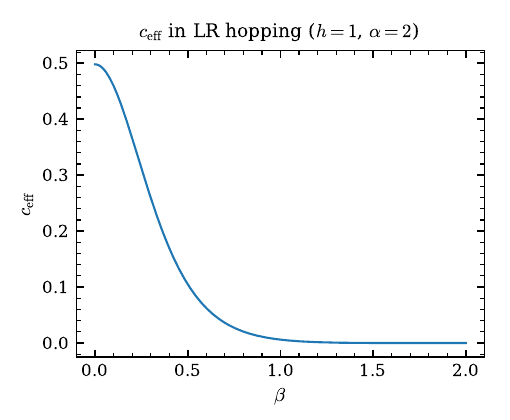}
\includegraphics[width=.47\textwidth]{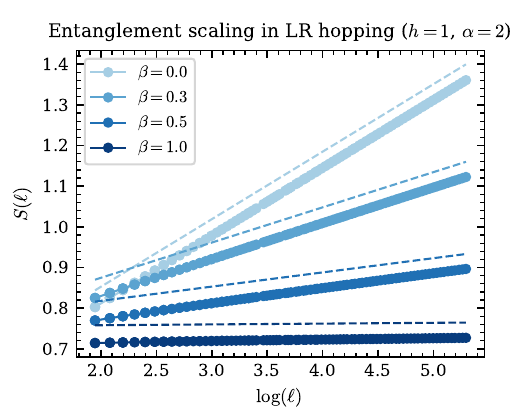}
\caption{\textbf{Entanglement entropy for $\alpha>1$ and initial critical state.} Entanglement entropy in the state~\eqref{eq:longrange} following a non-unitary evolution with long-range hopping in eq.~\eqref{eq:Hev}, $\alpha=2$, $\mu = 0$, and initial state with $h=1= \gamma$ in Eq.~\eqref{eq:init}. The left panel shows the analytical behavior of the effective central charge~\eqref{genEntropy} after the non-unitary evolution. In the right panel, the data points were obtained using Eq.~\eqref{eq:numericsent} for $\beta \in [0,1]$.  For all $\beta \neq 0$ shown, we obtain an altered effective central charge that agrees well with the analytical predictions in the left panel (dashed lines).}
\label{fig:entropylarger1}
\end{figure}

\paragraph{Long-range $\alpha>1$:} 
For $\alpha>1$, the symbol of the projection state is smooth and the non-unitary evolution does not generate any discontinuity, and if we start from a gapped ground state, the entanglement entropy remains area-law for all $\beta$. The only nontrivial case occurs at the critical point $h=1$, where the initial state already exhibits a discontinuity, $u_0(0)=1/\sqrt{2}$, $v_{0^{+}}(0)= 1/\sqrt{2}$ and $v_{2\pi^{-}}(0)= -1/\sqrt{2}$. This jump is responsible for the well-known $1/6$ coefficient governing the logarithmic growth of entanglement~\cite{calabrese2004,JinKorepin2004}. Evolving the state according to Eq.~\eqref{eq:longrange}, the magnitude of this discontinuity is progressively reduced, with $|g(\beta)|=(\cosh(2\beta G_{k=0}))^{-1}$. The corresponding logarithmic growth coefficient is then obtained by inserting this expression into Eq.~\eqref{genEntropy}. In Fig.~\ref{fig:entropylarger1} we show the analytical behavior of the effective central charge as a function of the measurement strength and we benchmark our prediction against exact numerics. In the limit $\beta\to \infty$, $c_{\rm eff}(\beta)\to 0$, which is compatible with the fact that the projection state obeys an area law behavior.

\paragraph{Long-range $\alpha=1$:} In the presence of long-range hopping in Eq.~\eqref{eq:Hev}, as soon as $\beta>0$, the system shows the typical features of the projection state, meaning area law and long-range correlations, as we prove in Appendix~\ref{app:symbol}.

\begin{figure}[t!]
\includegraphics[width=.47\textwidth]{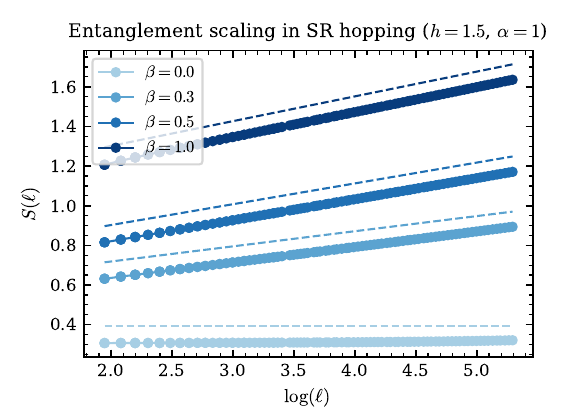}
\includegraphics[width=.47\textwidth]{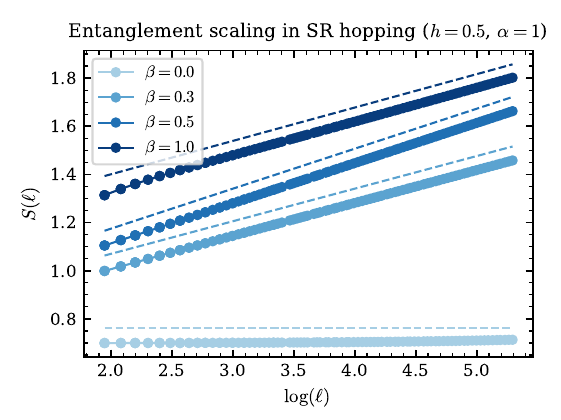}
\caption{\textbf{Entanglement entropy for $\alpha=1$.} Entanglement entropy in the state~\eqref{eq:longrange} following a non-unitary evolution with short-range hopping in eq.~\eqref{eq:Hev}, $\alpha=1$, $\mu = 0$, and initial state with $ \gamma=1$ and $h=1.5$ (left panel) and $h=0.5$ (right panel) in Eq.~\eqref{eq:init}.The data points were obtained using Eq.~\eqref{eq:numericsent} for $\beta \in [0,1]$.  For all $\beta \neq 0$ shown, we obtain an altered effective central charge that agrees well with the analytical predictions obtained by plugging Eq.~\eqref{eq:gentropy} into Eq.~\eqref{genEntropy} (dashed lines).}
\label{fig:entropy1}
\end{figure}

However, for short-range hopping, we expect the logarithmic growth coefficient of the entanglement to turn on continuously as $\beta>0$ and the factors $g(\beta)$ which enter in the effective central charge~\eqref{genEntropy} are 

\begin{equation}\label{eq:gentropy}
 \begin{split}
  h>1,\quad &g(\beta)=-2i\frac{\left(\cosh(\beta\lambda_0)+\frac{t-\mu}{\lambda_{0}}\sinh(\beta\lambda_{0})\right)\left(\frac{\pi}{2\lambda_{0}}\sinh(\beta\lambda_{0})\right)}{\left(\cosh(\beta\lambda_{0})+\frac{t-\mu}{\lambda_{0}}\sinh(\beta\lambda_{0})\right)^2+\left(\frac{\pi}{2\lambda_{0}}\sinh(\beta\lambda_{0})\right)^2}, \\
    h<1, \quad &g(\beta)=-2i\frac{\left(\cosh(\beta\lambda_{0})-\frac{t-\mu}{\lambda_{0}}\sinh(\beta\lambda_{0})\right)\left(\frac{\pi}{2\lambda_{0}}\sinh(\beta\lambda_{0})\right)}{\left(\cosh(\beta\lambda_{0})-\frac{t-\mu}{\lambda_{0}}\sinh(\beta\lambda_{0})\right)^2+\left(\frac{\pi}{2\lambda_{0}}\sinh(\beta\lambda_{0})\right)^2}, 
 \end{split} 
 \end{equation}
with $\lambda_0= \sqrt{(t-\mu)^2+\frac{\pi^2}{4}}$. We support our predictions in Fig.~\ref{fig:entropy1}. By contrast, at $h=1$, the initial state already exhibits the maximal logarithmic entanglement growth, with coefficient $1/6$. As $\beta>0$, this coefficient is expected to decrease smoothly with $\beta$, eventually saturating at the value associated with the projection state of $H$ in Eq.~\eqref{eq:Hev}. The factor $g(\beta)$ is in this case
\begin{equation}
g(\beta)=-2i\frac{\left(\cosh(\beta\lambda_{0})+\left(\frac{\pi}{2\lambda_0}+\frac{t-\mu}{\lambda_{0}}\right)\sinh(\beta\lambda_{0})\right)\left(\cosh(\beta\lambda_{0})+\left(\frac{\pi}{2\lambda_0}-\frac{t-\mu}{\lambda_{0}}\right)\sinh(\beta\lambda_{0})\right)}{\left(\cosh(\beta\lambda_{0})+\left(\frac{\pi}{2\lambda_0}+\frac{t-\mu}{\lambda_{0}}\right)\sinh(\beta\lambda_{0})\right)^2+\left(\cosh(\beta\lambda_{0})+\left(\frac{\pi}{2\lambda_0}-\frac{t-\mu}{\lambda_{0}}\right)\sinh(\beta\lambda_{0})\right)^2}. 
\end{equation}

\paragraph{Long-range regime $\alpha<1$:} When $\alpha<1$, the symbol of the projection state exhibits a finite discontinuity at $k=0$. In the presence of long-range hopping, the discontinuity is given by
\begin{equation}\label{eq:discalpha}
    |g| = \cos\!\left(\frac{\pi}{2}\alpha\right).
\end{equation}
This produces a logarithmic entanglement growth as soon as $\beta>0$. The corresponding coefficient increases as \(\alpha\) decreases and approaches $1/6$ in the limit \(\alpha\to 0\). When the evolution Hamiltonian~\eqref{eq:Hev} has short-range hopping, the projection state instead exhibits a discontinuity \( |g|=1 \), for all values of \(\alpha\). In this case, the evolution produces an instantaneous logarithmic growth coefficient of $1/6$ given by Eq.~\eqref{genEntropy}, identical to that of the critical Ising model.

\begin{figure}[t!]
\includegraphics[width=.47\textwidth]{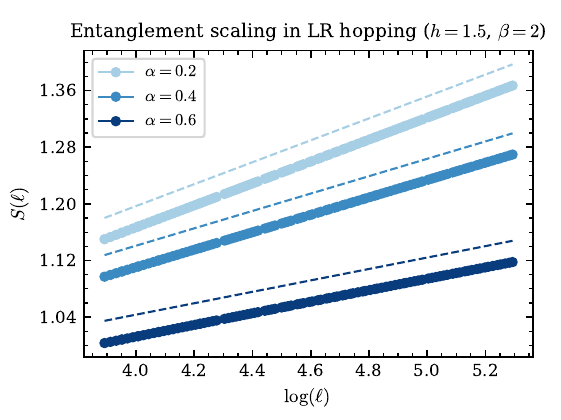}
\includegraphics[width=.47\textwidth]{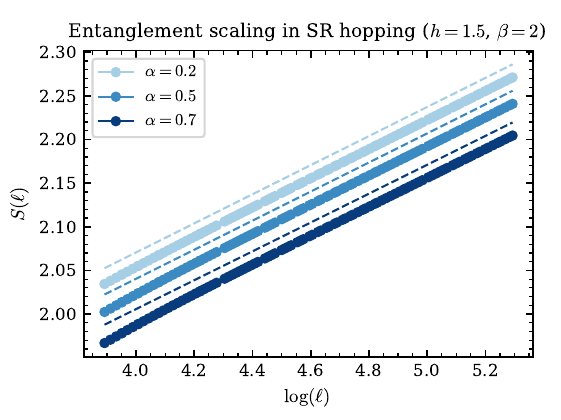}
\caption{\textbf{Entanglement entropy for $\alpha<1$.} Entanglement entropy in the state~\eqref{eq:longrange} following the non-unitary evolution in eq.~\eqref{eq:Hev}, with $\beta=2$, $\mu = 0$, and initial state with $ \gamma=1$ and $h=1.5$ in Eq.~\eqref{eq:init}. The data points were obtained using Eq.~\eqref{eq:numericsent} for $\alpha<1$.  For all $\alpha > 0$ shown, we obtain an altered logarithmic scaling of the entanglement entropy that agrees well with the analytical predictions obtained by plugging Eq.~\eqref{eq:discalpha} into Eq.~\eqref{genEntropy} (dashed lines left plot), or the pristine critical scaling effective central charge $c_{\rm eff}=1/2$ (dashed lines right plot).}
\end{figure}

\section{Correlation functions of the deformed wave function}\label{sec:correlations}
\label{sec:4}

In addition to the entanglement entropy, correlation functions also provide important insights into the structure and phases of quantum many-body systems. 
These correlators offer an independent and complementary diagnostic to the entanglement measures discussed above. Motivated by the Jordan-Wigner transformation relating the Majorana chain to the transverse field Ising model, we look at order parameter correlations given by \(\langle \sigma_i^x \sigma_j^x \rangle\), and its Kramers-Wannier dual---the disorder parameter correlations, given by the string correlator $ \langle \sigma_i^z \sigma_{i+1}^z \cdots \sigma_{j-1}^z\rangle$. Here $\sigma_i^x$ and $\sigma_i^z$ denote Pauli spin operators. We expect that in the ferromagnetically ordered (topological superconductor in the Majorana language) phase, we have long-range order in the order parameter and decaying disorder parameters, and vice versa in the disordered (trivial) phase.

Consider the order parameter correlator
\begin{equation} \label{eq:x-corr}
\rho_{n}^x = \bra{\psi(\beta)} \sigma_l^x \sigma_{l+n}^x \ket{\psi(\beta)}.
\end{equation}
Performing a Jordan-Wigner transformation, the spin operators can be expressed in terms of fermionic creation and annihilation operators. Introducing the Majorana-like operators
\begin{equation}
A_l = c_l + c_l^\dagger, 
\qquad 
B_l = c_l^\dagger - c_l ,
\end{equation}
the spin correlation functions can be rewritten as expectation values involving products of \(A_l\) and \(B_l\). Since the state $\ket{\psi(\beta)}$ is Gaussian, by applying the anticommutation relations satisfied by the operators $A_l$ and $B_l$, together with Wick’s theorem, these expectation values reduce to combinations of two-point correlators. Noting that the $\langle A_j A_m \rangle$ and $\langle B_j B_m \rangle$ correlators vanish due to sublattice symmetry, we define
\begin{equation}
H(m-j) = \langle \psi(\beta)| B_j A_m| \psi(\beta) \rangle ,
\end{equation}
and denoting the separation by \(n = m-j\), the \(x\)-spin correlation functions can be written compactly as determinants of Toeplitz matrices:
{\small
\begin{align}\label{eq:xxcorr}
\rho^{x}_{n} &=
\begin{vmatrix}
H(1) & H(2) & \cdots & H(n) \\
H(0) & H(1) &        & \vdots \\
\vdots &        & \ddots & H(2) \\
H(2-n) & \cdots & H(0)  & H(1)
\end{vmatrix}.
\end{align}
}

The disorder parameter correlation function given by the string correlator of Pauli-$Z$ operators can be transformed into the fermionic language as
\begin{align}
\nonumber
\rho_n^{\mathrm{str}} 
&= \langle \psi(\beta)| \sigma_l^z \sigma_{l+1}^z \cdots \sigma_{l+n-1}^z |\psi(\beta)\rangle \\
&= \langle \psi(\beta)|A_l B_l A_{l+1} B_{l+1} \cdots A_{l+n-1} B_{l+n-1} |\psi(\beta)\rangle .
\end{align}
Applying Wick’s theorem once again, the string correlator can be expressed as a determinant,
\begin{equation}\label{eq:str}
\rho_n^{\mathrm{str}}
= (-1)^n
\begin{vmatrix}
H(0) & H(1) & H(2) & \cdots & H(n-1) \\
H(-1) & H(0) & H(1)  & \cdots & H(n-2) \\
H(-2) & H(-1) & H(0)  & \cdots & H(n-3) \\
\vdots & \vdots & \vdots & \ddots & \vdots \\
H(1-n) & H(2-n) & H(3-n) & \cdots & H(0)
\end{vmatrix}.
\end{equation}
We note that the order and disorder correlators differ by a `half-site' Majorana translation, manifesting the Kramers-Wannier duality. These nonlocal string correlations in the fermionic variables are particularly sensitive to hidden order and topological features, making them a valuable complement to both local spin correlators and entanglement-based measures.
In the following, we focus on the absolute value of the correlators, neglecting possible oscillatory factors.

We want to study the asymptotic behavior of the determinant of the matrices above, and we observe that their entries can be written as
\begin{equation}\label{HH}
    H(n) = -\int_{0}^{2\pi} \frac{dk}{2\pi} e^{-i k n} (n(k,\beta)+g(k,\beta))
\end{equation}
We see that the string correlator has symbol $\mathcal{G}^{\mathrm{str}}(k,\beta)=n(k,\beta)+g(k,\beta)$, while the $x$-correlator has the shifted symbol $\mathcal{G}^x(k,\beta)= -(n(k,\beta)+g(k,\beta))e^{-ik}$.
We now state the standard results about the determinants of Toeplitz matrices generated by such symbols, where $\mathcal{G}(k,\beta)$ denotes a generic scalar symbol.
When a symbol $\mathcal{G}(k,\beta)$ is continuous and has zero winding number, we can apply the \textit{Szegő theorem}, which yields the following asymptotic formula
\begin{equation}\label{eq:scalarsymbolcont}
    \log [\det T_{\ell}(\mathcal{G})]
= \frac{\ell}{2\pi} \int_{0}^{2\pi} \log[\mathcal{G}(k,\beta)]dk  + O(1).
\end{equation}
Since $\mathcal{G}(k,\beta)$ takes values on the unit circle, $\log \mathcal{G}(k,\beta)$ is purely imaginary, so the integral contributes only a phase and does not affect $|\det T_\ell(\mathcal{G})|$  . 
As a result, when the symbol is smooth and has zero winding number, the determinant, and therefore the correlator, saturates to a nonzero constant.

To handle symbols with nonzero winding number, we must go beyond the standard Szegő formula. One can analyze the determinant by extending the symbol into the complex plane. The asymptotic behavior is then dominated by the singularity closest to the unit circle, which directly sets the correlation length. As a result, the leading contribution decays exponentially~\cite{fisher1968toeplitz,HartwigFisher1969,Krasovsky2011,ares2021full}. 

Finally, when the symbol is discontinuous, it can be factorized as
\begin{equation}
\mathcal{G}(k,\beta)=V(k,\beta)\prod_{\sigma=1}^{r}e^{it_{\sigma}(k-k_{\sigma}-\pi\text{sign}(k-k_{\sigma}))},\label{factorizationdisc}
\end{equation}
such that $V(k,\beta)$ is continuous and with zero winding number. The Fisher–Hartwig theorem~\cite{Its2005,DeiftItsKrasovsky2011} for a scalar symbol $\mathcal{G}(k,\beta)$, similarly to eq.~\eqref{eq:conjecture1}, reads
\begin{equation}\label{eq:scalarsymbol}
    \log [\det T_{\ell}(\mathcal{G})]
= \frac{\ell}{2\pi} \int_{0}^{2\pi} \log[V(k,\beta)]dk  -\log\ell \sum_{\sigma=1}^r t^2_{\sigma} + O(1).
\end{equation}
We now have all the tools to study the behavior of the different correlation functions in three different regimes.

\begin{figure}[t!]
\includegraphics[width=.47\textwidth]{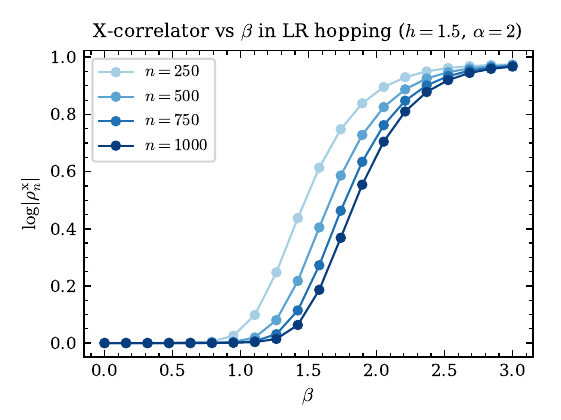}
\includegraphics[width=.47\textwidth]{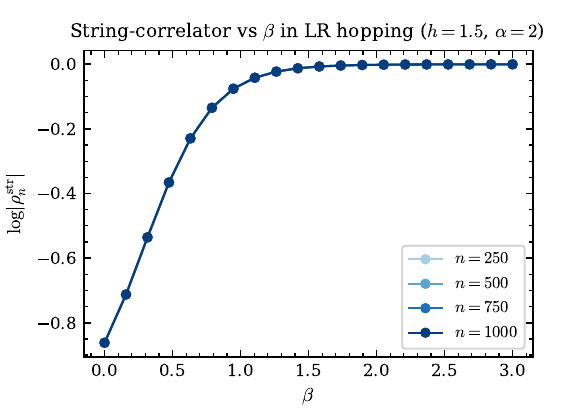}
\caption{\textbf{Correlation functions vs measurement strength $\beta$ for $\alpha>1$.} Correlations in the state~\eqref{eq:longrange} following a non-unitary evolution with long-range hopping in eq.~\eqref{eq:Hev}, $\alpha=2$ and initial state with $h=1.5, \gamma=1$ in Eq.~\eqref{eq:init}. Data points were obtained using Eqs.~\eqref{eq:xxcorr} and~\eqref{eq:str} for different system size. The fact that increasing the distance between two sites $i$ and $i+\ell$ shifts the curve for $\rho^x_{i,i+\ell}$ towards larger values of $\beta$ is consistent with the absence of a phase transition at finite $\beta$.}
\label{fig:alphalarger1corr2}
\end{figure}

In the following discussion, we usually refer to a single symbol even though we have both the $x$- and string-correlator symbols. This is because they are related by a continuous factor: they are either both continuous or both discontinuous, and since the factor $e^{-ik}$ shifts the winding number by $-1$, the winding number of one symbol determines that of the other. With this in mind, we define the ``reference'' symbol $\mathcal{G}(k,\beta) \equiv \mathcal{G}^{\mathrm{str}}(k,\beta)$, since the string symbol has the simpler form, and we restrict the analysis to this one (see Appendix~\ref{app:Hev} and Appendix~\ref{app:symbol}) — the properties of the $x$-correlator symbol then follow. Note also that the entries of the block symbol in Eq.~\eqref{eq:corr} are directly related to the real and imaginary parts of $\mathcal{G}(k,\beta)$\footnote{This happens because $u_k(\beta)$ and $v_k(\beta)$ remain real for our choice of evolution Hamiltonians.}. 
\paragraph{Long-range regime $\alpha>1$:} For $\alpha>1$, the symbol $\mathcal{G}(k,\beta)$ evolves continuously as a function of $k$ and no additional discontinuities are generated during the evolution. In the limit $\beta\to \infty$, the evolution operator $e^{\beta H}$ becomes a projector onto the highest-energy eigenstate of the evolution Hamiltonian. As shown in App.~\ref{app:Hev}, for the Hamiltonians~\eqref{eq:Hev}, this projected state is topological (ordered) when $\alpha>1$: the $x$-correlator will saturate to a non-zero value, while the string-correlator exponentially decays. Consequently, the large-$\beta$ limit always corresponds to a topological phase. For $h<1$ the initial state is already topological, and no phase transition is expected during the evolution. At the critical point $h=1$, we observe a crossover to a critical-like regime at non-zero $\beta$, eventually realizing a topological phase only in the limit $\beta\to \infty$. As we tune $\beta$, the critical-like regime exhibits a continuous evolution of the power-law decay exponents, with the exponent of the 
$x$-correlator approaching zero as $\beta\to\infty$, 
\begin{equation}\label{eq:betacorr}
    \rho_{lj}^x\sim \frac{1}{|l-j|^{2\Delta(\beta)}}, \quad \Delta(\beta)=\frac{1}{2}\left(\frac{1}{\pi}\arg \left[-\text{sgn}(G_{0^+}) \tanh(2\beta\lambda_{0^+})-i\sech(2\beta\lambda_{0^+})\right]+1\right)^2
\end{equation}
where $\lambda_{0^+}=\sqrt{G_{0^+}^2+F_{0^+}^2}$.
\begin{figure}[t!]
\includegraphics[width=.47\textwidth]{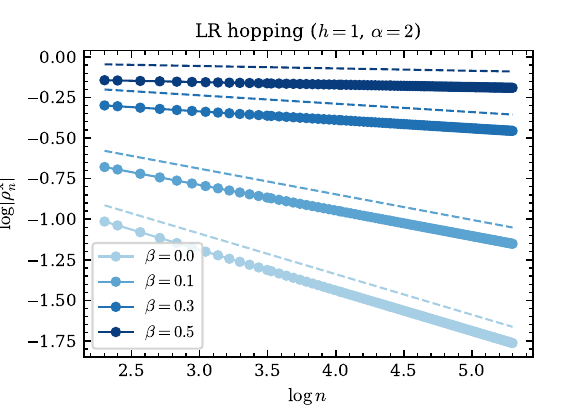}
\includegraphics[width=.47\textwidth]{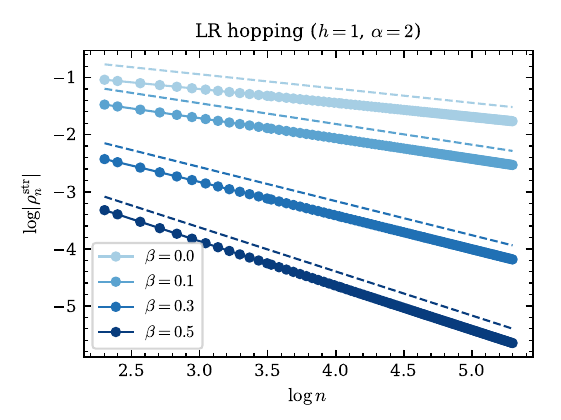}
\caption{\textbf{Correlation functions for $\alpha>1$.} Correlations in the state~\eqref{eq:longrange} following a non-unitary evolution with long-range hopping in eq.~\eqref{eq:Hev}, $\alpha=2$ and initial critical state, $h=\gamma=1$ in Eq.~\eqref{eq:init}. Data points were obtained using Eqs.~\eqref{eq:xxcorr} and~\eqref{eq:str} for $\beta \in [0,0.5]$.  For all $\beta \neq 0$ shown, we obtain altered power-law correlations with exponents that agree well with the analytical predictions in Eqs.~\eqref{eq:betacorr} and \eqref{eq:betacorrstr} (dashed lines).}
\label{fig:alphalarger1corr}
\end{figure}
This behavior is consistent with the fact that the projection state is topological, implying saturation of the 
$x$-correlator at long distances. We cross-check this result against exact numerical computations in the left panel of Fig.~\ref{fig:alphalarger1corr}. In the right panel, we also cross-check the analytical behavior of the string correlator,
\begin{equation}\label{eq:betacorrstr}
    \rho_{lj}^{\mathrm{str}}\sim \frac{1}{|l-j|^{2\Delta(\beta)}}, \quad \Delta(\beta)=\frac{1}{2}\left(\frac{1}{\pi}\arg \left[-\text{sgn}(G_{0^+}) \tanh(2\beta\lambda_{0^+})-i\sech(2\beta\lambda_{0^+})\right]\right)^2
\end{equation}
We observe that the power-law exponent increases with the measurement strength, implying that in the $\beta \to \infty$ limit the string correlator vanishes, consistent with the system flowing to a topological phase. For $h>1$, the unevolved state is trivial, while the projection state is topological. Therefore, a transition from a trivial to a topological phase must occur at some value of $\beta$. This transition can be characterized by the winding number of the 
string correlator, which must change from zero to one. However, since the symbol has unit modulus for all values of $\beta$, such a change cannot occur through any finite, continuous deformation. As a result, the transition is pushed to $\beta\to \infty$. This conclusion is supported by examining the two-point correlation function at fixed distance $\ell$ as a function of $\beta$, as shown in Fig.~\ref{fig:alphalarger1corr2}: the apparent saturation to a constant value shifts to larger $\beta$ as the distance increases, indicating that in the thermodynamic limit the transition indeed occurs only at infinite $\beta$. The discussion above holds both in the presence of short-range or long-range hopping terms in the evolution Hamiltonian~\eqref{eq:Hev}.

\begin{figure}[t!]
\includegraphics[width=.47\textwidth]{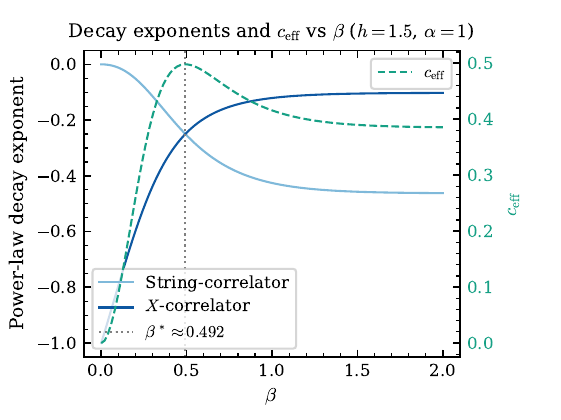}
\includegraphics[width=.47\textwidth]{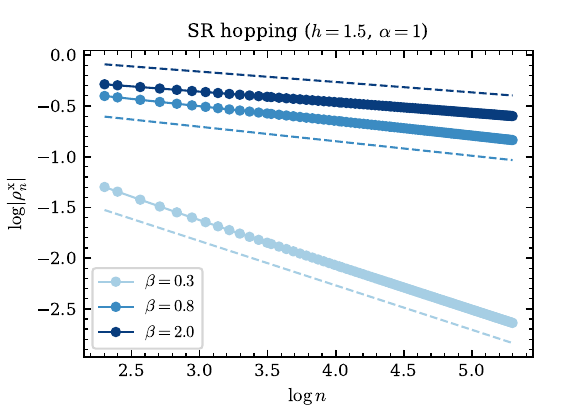}
\caption{\textbf{Correlation functions for $\alpha=1$.} Correlations in the state~\eqref{eq:longrange} following a non-unitary evolution with short-range hopping in eq.~\eqref{eq:Hev}, $t=-1$, $\alpha=1$ and initial  state with $h=1.5, \gamma=1$ in Eq.~\eqref{eq:init}. The left panel shows the analytical behavior of the power-law exponent of the $x$ and string-correlators after the non-unitary evolution. For large values of the measurement strength, they tend to the one of the evolution Hamiltonian. In the right panel, the data points were obtained using Eq.~\eqref{eq:xxcorr} for $\beta \in [0,2]$.  For all $\beta \neq 0$ shown, we obtain altered power-law correlations with exponents that agree well with the analytical predictions in the left panel (dashed lines). }
\label{fig:alpha1corr}
\end{figure}

\paragraph{Long-range regime $\alpha=1$:} When $\alpha=1$ the evolution is no longer continuous in $k$. If the evolution~\eqref{eq:Hev} is characterized by long-range hopping, then the state enters into a topological phase as soon as $\beta>0$ (see Appendix~\ref{app:symbol}). However, in the presence of short-range hopping in Eq.~\eqref{eq:Hev}, by analyzing the case $h>1$ (the others are similar), we notice that the 
$x$ and string correlation functions exhibit power-law decay with exponents that evolve monotonically with 
$\beta$, converging to those of the projection state (which is not in a topological phase for $\alpha=1$), as confirmed by the numerical plots in Fig.~\ref{fig:alpha1corr}. This implies that there is not a transition from a trivial to a topological phase, not even at $\beta\to\infty$.

We observe a curious coincidence in the first panel of Fig.~\ref{fig:alpha1corr}. Overlaying the plot of the correlation exponent with the coefficient of the logarithmic decay of entanglement entropy, we find that the exponents cross each other at $-1/4$, while the effective central charge saturates to a value of $1/2$ at the same value of $\beta$. These exponents and effective central charge match exactly with the Ising conformal field theory (CFT), which is realized by the transverse field Ising model at criticality---dual to the critical Majorana chain. We explain this coincidence within the framework of fermionic Gaussian states, and connect it to an emergent Kramers-Wannier (KW) symmetry which highly constrains the allowed long-distance behavior.

We know that the long-distance properties of the Gaussian state is dictated by the non-analytic features of the symbol in momentum space. As the KW duality of the spin-chain acts as a Majorana translation operator of the fermion chain~\cite{Seiberg2024}, the symbol transforms under KW as 
\begin{equation}
    n'(k,\beta)=-n(k,\beta)\cos(k)+i g(k,\beta)\sin(k), \quad g'(k,\beta)= g(k,\beta)\cos(k)-i n(k,\beta)\sin(k).
\end{equation}
We define an emergent KW symmetric state as a Gaussian state which satisfies $n'(k) = n(k), g'(k) = g(k)$ as we take the limits $k \to {k^\ast}^{\pm}$, where $k^\ast$ are the discontinuities of the symbol. In our case, we only have one discontinuity ($k^\ast = 0$) and we find that enforcing the emergent KW symmetry condition at this point constrains the state to have the same discontinuity properties as the critical Majorana chain, which fixes both its correlation decay exponents to be $-1/4$ and the effective central charge to be $1/2$. For our particular deformation, we perform a calculation to find the emergent KW symmetry exists at $\beta^* \approx 0.492$, in agreement with the plot (see Appendix~\ref{App:KW} for details).

\begin{figure}[t!]
\includegraphics[width=.47\textwidth]{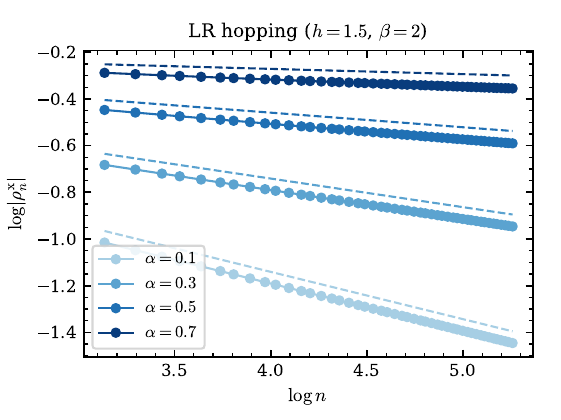}
\includegraphics[width=.47\textwidth]{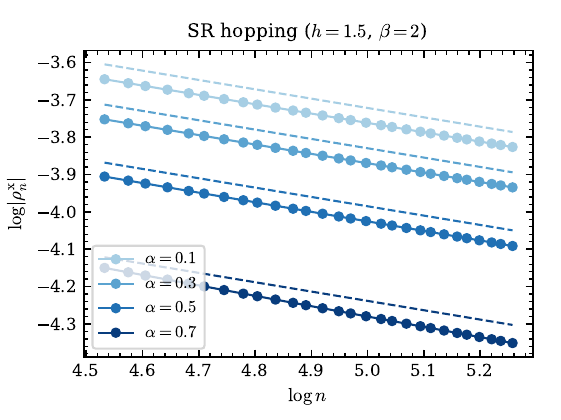}
\caption{\textbf{Correlation functions for $\alpha<1$.} Correlations in the state~\eqref{eq:longrange} following a non-unitary evolution with long-range (left panel) or short-range (right panel) hopping in eq.~\eqref{eq:Hev}, $\beta=2$ and initial state with $h=1.5, \gamma=1$ in Eq.~\eqref{eq:init}. Data points were obtained using Eq.~\eqref{eq:xxcorr} for $\alpha \in [0.1,0.7]$.  For all values of $\alpha$ shown, we obtain altered power-law correlations with exponents that continuously change as a function of $\alpha$ in the left panel, as predicted by Eq.~\eqref{eq:rhox} shown by the dashed lines. For short-range hopping, the power-law decays reduce to the Ising critical exponent $\rho^x_{|i-j|}\sim |i-j|^{-1/4}$, i.e. $\Delta(\alpha)=1/8$ (dashed lines).}
\label{fig:alphasmaller1corr}
\end{figure}

\paragraph{Long-range regime $\alpha<1$:} For 
$\alpha<1$, an even more striking phenomenon occurs: for any 
$\beta>0$, the correlations instantaneously acquire the power-law decay of the projection state, similarly to the instantaneous logarithmic growth of the entanglement discussed in Section~\ref{sec:entanglement}. The corresponding decay exponents can be computed using Eq.~\eqref{eq:scalarsymbol}. In the short-range hopping case, these exponents coincide with those of the critical Ising model and are independent of $\alpha$. Indeed, the symbol exhibits a discontinuity at $k=0$ and the goal is to choose $t_1$ in Eq.~\eqref{factorizationdisc} such that this discontinuity is absorbed into the Fisher–Hartwig factor, leaving a smooth residual symbol with zero winding number. This requirement is crucial, since a nonzero winding number would lead to an exponential decay rather than to a power-law behavior. Because the discontinuity corresponds to a phase jump of $\pi$, it can be compensated by choosing $t_1$ to be a half-integer. The remaining freedom is fixed by the direction in which the unfactorized symbol winds around the origin. For the string correlator, the symbol has zero winding number and it goes around the origin in the clockwise direction. Therefore, $t_1$ must be equal to $1/2$. By the same reasoning, the symbol associated with the 
$x$-correlator winds in the opposite direction, leading instead to $t_1=-1/2$. By applying Eq.~\eqref{eq:scalarsymbol}, we find $\rho^x_{n}\sim n^{-1/4}$ and $\rho_n^s\sim n^{-1/4}$.
In the long-range hopping case 
\begin{equation}\label{eq:rhox}
\rho_{lj}^x\sim \frac{1}{|l-j|^{2\Delta(\alpha)}}, \quad \Delta(\alpha)=\frac{1}{2}\left(\frac{\alpha-1}{2}\right)^2,
\end{equation}
what power-law exponent depending on $\alpha$ and approaching the Ising values as $\alpha\to 0$. This behavior is corroborated by the numerical results shown in Fig.~\ref{fig:alphasmaller1corr} for the long-range hopping (left panel) and short-range hopping case.

\section{Entanglement spectrum and partition function}\label{sec:spectrum}

\begin{figure}[t!]
\includegraphics[width=.47\textwidth]{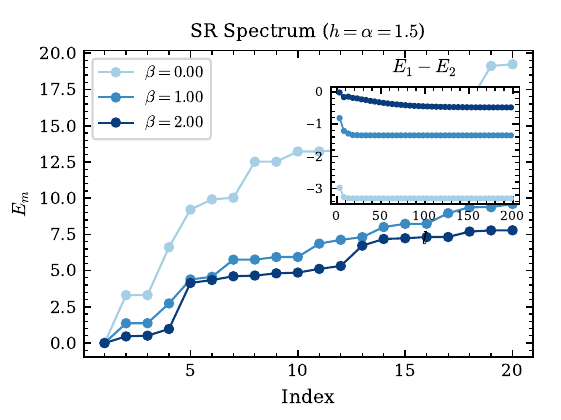}
\includegraphics[width=.47\textwidth]{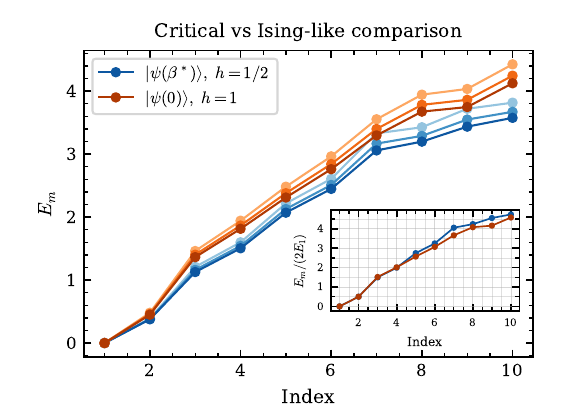}
\caption{\textbf{Entanglement spectrum} Left panel: entanglement spectrum for different values of $\beta$, $\alpha=1.5$, starting from the initial state with $h=1.5,\gamma=1$. Although the lowest-lying levels may appear nearly degenerate at first sight, the inset shows the difference between the first two eigenvalues of $\rho_A$ as a function of system size. Its saturation to a finite nonzero value demonstrates that these levels are not truly degenerate, so at finite $\beta$ we are not in a topological phase. The right panel shows the comparison between the spectrum of $\rho_A$ at $\beta=0$, with the initial state tuned to the critical point $\gamma=h=1$, and the spectrum at $\beta=\beta^*$ and $\alpha=1$, where the long-distance behavior is consistent with critical Ising scaling. The different shades correspond to subsystem sizes $\ell=500, 1000,2000$. The spectra deviate at larger values of the eigenvalue $E_m$. 
It is unclear whether the entanglement spectrum for $\ket{\psi(\beta^\ast)}$ is truly distinct from the universal predictions of Ising CFT, or whether the deviations we observe are finite size effects.} 
\label{fig:alphalarger1corr2sp}
\end{figure}
Following the approach introduced in studies of the entanglement spectrum of topological insulators and superconductors (see Ref.~\cite{Fidkowski2010}), the reduced density matrix of a Gaussian fermionic state can be interpreted as a thermal state of an effective entanglement Hamiltonian. This representation provides access to the full entanglement spectrum, which contains information beyond entanglement entropies and can reveal universal and topological features of the state~\cite{LiHaldane2008,Pollmann2010}. For this reason, we study it here to corroborate the absence of finite-$\beta$ phase transitions. 
Let $\Gamma$ be the correlation matrix restricted to subsystem $A$, with eigenvalues \( \{\pm \lambda_r\}_{r=1}^{\ell} \) with \( \lambda_r \in [-1,1] \). Each eigenvalue defines a single-particle entanglement energy
\begin{equation}
\epsilon_r \;=\; 2\,\mathrm{arctanh}(\lambda_r)
\;=\; \ln\!\left(\frac{1+\lambda_r}{1-\lambda_r}\right).
\label{eq:eps-from-lambda}
\end{equation}
In terms of these modes, the reduced density matrix $\rho_A$ factorizes as
\begin{equation}
\rho_A \;= \frac{1}{Z}e^{-\sum_r \epsilon_rd_r^\dagger d_r}=\; \bigotimes_{r=1}^{\ell} \rho_{A,r},
\end{equation}
with $\rho_{A,r} = (1+e^{-\epsilon_r})^{-1} e^{-\epsilon_r d_r^\dagger d_r}$. 
The many-body eigenvalues of $\rho_A$ are obtained by independently occupying or emptying each single-particle entanglement mode,
\begin{equation}
p_{\{s_r\}}
= \prod_{r=1}^{\ell} \frac{1+s_r\lambda_r}{2},
\qquad s_r=\pm1,
\label{eq:manybody-probs}
\end{equation}
where $s_r=-1$ ($s_r=+1$) corresponds to an occupied (empty) entanglement mode. Equivalently, the same eigenvalues can be written in the thermal form
\begin{equation}
p_m=\frac{e^{-E_m}}{Z},
\end{equation}
where $E_m$ are the eigenvalues of the entanglement Hamiltonian
\begin{equation}
H_E=\sum_r \epsilon_r\, d_r^\dagger d_r,
\end{equation}
with single-particle entanglement energies $\epsilon_r$. The set of levels $\{E_m\}$ defines the \emph{entanglement spectrum}, which contains much finer information than the entanglement entropy alone, as it resolves the full operator content of the reduced density matrix. At criticality, the low-lying part of the entanglement spectrum is expected to be described by boundary conformal field theory: the entanglement Hamiltonian can be mapped to the generator of translations along an effective strip, so that its universal structure is organized into conformal towers determined by the boundary conditions at the entangling cut~\cite{CardyTonni2016}.

To give a concrete example, let us take the ground state of Eq.~\eqref{eq:HinitF} with $\gamma=1$: the phase with $h<1$ is topological, in the sense that it is characterized by having unpaired Majorana modes at the boundary. This means that in a chain of length $L\gg\xi$, with $\xi$ the correlation length, there are two ground states, degenerate up to a splitting $e^{-L/\xi}$. The entanglement spectrum also contains a signature of the topological phase since the multiplicity of all eigenvalues in the Schmidt
decomposition is doubled. In this sense, the degeneracy of the entanglement spectrum can also be considered as a probe of the presence of a topological phase. 

In the previous section, we argued that a phase transition between the trivial ($h\geq 1$) and topological phase occurs only for $\alpha> 1$ (for $\alpha=1$, this occurs only for evolution Hamiltonians~\eqref{eq:Hev} with long-range hopping) and $\beta\to\infty$. If this picture is correct, then for $h\geq1$ and $\alpha>1$ we should not observe any twofold degeneracy in the entanglement spectrum at finite $\beta$. Although the eigenvalues shown in Fig.~\ref{fig:alphalarger1corr2sp} appear very close to one another, a more detailed analysis shows that they are not degenerate. In particular, the gap between the first two excited entanglement levels saturates to a finite value as the system size increases, rather than closing. This supports the absence of degeneracy and hence the absence of a topological phase in this regime. In the same figure, we also compare the low-lying entanglement spectra of the critical Ising model and of the state~\eqref{eq:Hev} at $\alpha=1$ and $\beta=\beta^*$. The two spectra are close, but not identical. The inset shows the spectra rescaled by the first gap, so that one can directly compare the universal level ratios. These ratios can be explained as follows: at criticality, the low-lying entanglement spectrum should organize into conformal towers determined by the boundary conditions~\cite{CardyTonni2016,Roy2020}, and deviations from the Ising values suggest a departure from genuine critical behavior. However, we cannot rule out whether the deviations in Fig.~\ref{fig:alphalarger1corr2sp} are also influenced by finite size effects.

\section{Conclusions}

We studied how a restricted class of long-range Gaussian deformations modify the structure of fermionic Gaussian states, mainly focusing on diagnostics like the entanglement entropy and the correlation functions. Our results show that the key quantity controlling the long-distance physics is the decay exponent of the long-range deformation, $\alpha$. For $\alpha>1$, the deformation remains effectively weak at large distances: finite deformation strength, $\beta$, produces only smooth crossovers, and when a topological regime is reached this occurs only asymptotically as $\beta\to \infty$. At $\alpha=1$, the system can show a logarithmic scaling of the entanglement and algebraically decaying correlations with continuously variable universal properties, even when initialized in a gapped phase. For $\alpha<1$, instead, the deformation becomes sufficiently extended to directly imprint a critical-like long-distance behavior at arbitrarily small $\beta$, leading to immediate logarithmic entanglement growth and power-law correlations.

These results provide a simple and analytically controlled example of how long-range deformations can generate new infrared behavior starting from an exactly solvable reference state. In particular, they show that signatures of criticality can emerge without tuning to a conventional equilibrium critical point, and instead arise directly from the spatial structure of the deformation itself. The agreement between analytical predictions and exact numerical calculations for both entanglement and correlation functions supports this picture. The entanglement spectrum further confirms the absence of a genuine finite-$\beta$ transition in the regimes where the flow is only asymptotic. We have also examined a boundary partition function as an additional probe of the transition; however, we do not present these results here, as they do not provide significant additional insight beyond the results already discussed. Indeed, this quantity plays a role analogous to an ordinary partition function, but with the initial state imposing boundary conditions rather than performing a trace over all states. From our analysis, this object remains smooth for all finite values of the deformation parameter. In particular, we do not observe any non-analyticity or other sharp feature that could signal a genuine phase transition at finite deformation strength. 

More broadly, our work suggests that long-range wave function deformations provide a nontrivial yet tractable class of models to discover new many-body physics that can arise when notions of strict locality are relaxed. An important direction for future work is to understand how much of this picture survives beyond the Gaussian setting, where interactions may qualitatively alter both the thermodynamic limit and the nature of the induced critical behavior~\cite{liu2026}. Another natural extension would be to explore whether similar deformations can be used to construct broader families of critical or topological states, and to clarify the role of emergent dualities and universal scaling structures in such non-unitary long-range flows. We have mentioned that for specific values of the parameters of our model, there is an emergent KW symmetry, which we observe by studying the long-distance properties of the system in terms of entanglement or correlation functions. We would like to understand better if such a notion of emergent KW symmetry exists in the presence of arbitrary (non-Gaussian)
perturbations. Can it impose similar non-trivial constraints at long distances? We leave these questions to future work. Finally, it would be interesting to understand whether such long-range deformations can be realized in practice, for instance, through digital platforms~\cite{FossFeig}, measurement protocols~\cite{murciano2023}, or dissipative dynamics~\cite{Lin2025}.

\section*{Acknowledgements}

We are especially grateful to Jason Alicea, whose original idea initiated this project, namely to study long-range non-unitary deformations of the Ising model in the context of measurement-altered criticality. We also thank Filiberto Ares, Maurizio Fagotti, Yue Liu, David Mross, Stephen Naus, Silvia Pappalardi, Guido Pupillo, Andrea Solfanelli, and Pablo Sala for useful discussions about long-range systems and deformations.

\bibliographystyle{unsrtnat}
\bibliography{references}

\appendix
\section{Emergent Kramers-Wannier Symmetry}
\label{App:KW}
Under a Kramers-Wannier (KW) transformation, the Majorana fermions simply get translated by one Majorana site in real space. In momentum space, this corresponds to $\gamma_{k,e}\mapsto \gamma_{k,o}$ and $\gamma_{k,o}\mapsto e^{ik}\gamma_{k,e}$. Here $\gamma_k^{e,o} \equiv \frac{e^{-i\frac{\pi}{4}}}{\sqrt{N}}\sum_je^{-ikj}\gamma_{j}^{e,o}$. We construct a matrix that encodes the KW transformation in the momentum space of the Majorana operators
\begin{equation}
S=\begin{pmatrix}
    0 & 1 \\
    e^{ik } & 0
\end{pmatrix}.
\end{equation}
To go from Majorana to complex fermions $(c_k,c_{2\pi-k}^\dagger)$, we apply the transformation 
\begin{equation}
    M=\begin{pmatrix}
        1&1\\
        i&-i
    \end{pmatrix}.
\end{equation}
If we combine both of them, we get 
\begin{equation}
    R=M^{-1}S M=e^{ik/2}\begin{pmatrix}
        \sin(k/2)&-i\cos(k/2)\\
        i\cos(k/2) &  -\sin(k/2)
    \end{pmatrix}.
\end{equation}
If we consider the symbol 
\begin{equation}
\begin{pmatrix}
    n(k,\beta)&  g(k,\beta)\\
     g^*(k,\beta)&-n(k,\beta)
\end{pmatrix},   
\end{equation}
where
\begin{equation}
n(k,\beta) = \frac{|u_k(\beta)|^2 - |v_k(\beta)|^2}{|u_k(\beta)|^2 + |v_k(\beta)|^2}, \quad
g(k,\beta) = \frac{-2 i u_k^*(\beta) v_k(\beta)}{|u_k(\beta)|^2 + |v_k(\beta)|^2},
\end{equation}
under KW we get 
\begin{equation}\label{eq:KW}
    n'(k,\beta)=-n(k,\beta)\cos(k)+i g(k,\beta)\sin(k), \quad g'(k,\beta)= g(k,\beta)\cos(k)-i n(k,\beta)\sin(k).
\end{equation}
For the ground state of the transverse field Ising model (Majorana chain) at the critical point, $n(k,0)=\frac{1-\cos(k)}{\sqrt{(1-\cos(k))^2+\sin(k)^2}}$, $ g(k,0)=\frac{-i\sin(k)}{\sqrt{(1-\cos(k))^2+\sin(k)^2}}$ and one can explicitly check that $n'(k,0)=n(k,0)$ and $g'(k,0)=g(k,0)$ for all momenta. For our purposes, we wish to study the symbol when we let the system evolve with a non-unitary Hamiltonian with short-range hopping and long-range pairing ($\alpha=1$, $t=-1$, $\mu = 0$).

It can be useful to write down the evolved $u_k(\beta)$ and $v_k(\beta)$:

\begin{equation}
    \begin{split}
        u_k(\beta)=&u_k(0) \cosh \left(\beta  \sqrt{f_{\alpha}^2+\cos(k)^2}\right)-\frac{\sinh \left(\beta  \sqrt{f_{\alpha}^2+\cos(k)^2}\right)
   (\cos(k) u_k(0)-f_{\alpha} v_k(0))}{\sqrt{f_{\alpha}^2+\cos(k)^2}}\\
   v_k(\beta)=& v_k(0) \cosh \left(\beta  \sqrt{f_{\alpha}^2+\cos(k)^2}\right)+\frac{\sinh \left(\beta  \sqrt{f_{\alpha}^2+\cos(k)^2}\right)
   (\cos(k) v_k(0)+f_{\alpha} u_k(0))}{\sqrt{f_{\alpha}^2+\cos(k)^2}}
    \end{split}
    \label{evolKW}
\end{equation}
with $f_{\alpha}=\mathrm{Im}[\mathrm{Li}_{\alpha}(e^{ik})]$ and 
\begin{equation}
    \begin{split}
    u_k(0)=&\frac{1}{\sqrt{2}}\sqrt{\frac{\sqrt{(h-\cos (k))^2+\gamma^2\sin ^2(k)}+h-\cos (k)}{\sqrt{(h-\cos (k))^2+\gamma^2\sin
   ^2(k)}}}, \\ 
   v_k(0)=&\frac{\gamma\sin (k)}{\sqrt{2} \sqrt[4]{(h-\cos (k))^2+\gamma^2\sin ^2(k)} \sqrt{\sqrt{(h-\cos (k))^2+\gamma^2\sin
   ^2(k)}+h-\cos (k)}}.
\end{split}
\end{equation}
which for $h>1$ and $k\approx0$ takes the values $u_{k\approx0}(0)=1$ and $v_{k\approx0}(0)=0$.
When $\alpha = 1$ and $k\approx 0$, $f_{1}=\mathrm{Im}[\mathrm{Li}_{\alpha=1}(e^{ik})]=k/2-\mathrm{sign}(k)\pi/2$. To find the critical value of $\beta$ that 'reproduces' the Ising criticality, we need to impose that 
\begin{equation}
    \frac{u_{0^{\pm}}(\beta)}{\sqrt{|u_{0^{\pm}}(\beta)|^2+|v_{0^{\pm}}(\beta)|^2}}=\frac{1}{\sqrt{2}},\qquad \frac{v_{0^{\pm}}(\beta)}{\sqrt{|u_{0^{\pm}}(\beta)|^2+|v_{0^{\pm}}(\beta)|^2}}=\frac{\pm 1}{\sqrt{2}},
\end{equation}
such that $g(0^{\pm},\beta)=\mp i$ and we have $c_{\mathrm{eff}}=\frac{1}{2}$. Using \eqref{evolKW},
\begin{equation}
    u_{0^{\pm}}(\beta)=\cosh \left(\frac{1}{2} \sqrt{4+\pi ^2} \beta \right)-\frac{2 \sinh \left(\frac{1}{2} \sqrt{4+\pi
   ^2} \beta \right)}{\sqrt{4+\pi ^2}}, \qquad v_{0^{\pm}}(\beta)=\pm\frac{\pi  \sinh \left(\frac{1}{2} \sqrt{4+\pi ^2} \beta \right)}{\sqrt{4+\pi ^2}}.
\end{equation}

The critical value of $\beta$ is 
\begin{equation}\beta^* 
= \frac{2}{\sqrt{4+\pi^2}}\,\operatorname{arctanh}\!\left(\frac{\sqrt{4+\pi^2}}{2+\pi}\right)
\approx 0.492.
\end{equation}

If we compute Eq. \eqref{eq:KW} and we plot them for an arbitrary value of $h$, $\beta^*(h)$ as a function of $k$, we observe that they are not equal to $n(k,\beta)$ and $g(k,\beta)$, respectively. This is consistent as the emergent KW symmetry condition should only depend on the long-distance physics, which is influenced purely by the discontinuity, and is therefore realized by taking the limit of the condition (\ref{eq:KW}) on either side of the discontinuity point in momentum space.

Similarly, we note that the emergent KW symmetry is present in the case where the trivial state is deformed by the short-range hopping, long-range pairing Hamiltonian with $\alpha < 1$, where we observed Ising exponents for correlation functions and an effective central charge of $1/2$ for all values of $\beta>0$. We conclude that emergent KW symmetry in free-fermion Gaussian states is an IR equivalence condition purely dependent on the features of the symbol near its discontinuities, such that the universal long-distance behavior remains invariant under a KW duality but not the short-distance physics, necessarily.

\section{Phases of evolution Hamiltonians}\label{app:Hev}

The evolution given by Eq.~\eqref{eq:longrange} becomes a projection onto the 
highest-energy eigenstate of $H$ in the limit $\beta \to \infty$. It is therefore 
natural to characterize the phase diagram of this state. Recalling 
Eq.~\eqref{diagH}, the highest-energy eigenstate $\ket{\mathrm{PS}}$ satisfies
\begin{equation}
    \psi_k^\dagger\ket{\mathrm{PS}} = 0,\quad \forall\, k.\label{esc}
\end{equation}
It admits a momentum-space decomposition of the form (we ignore the details about $k=0,\pi$, since eventually we are interested in the thermodynamic limit) 
\begin{equation}
\ket{\mathrm{PS}}=\bigotimes_k\ket{\mathrm{PS}_k}=\prod_k\left(u_k^\mathrm{PS}+v_k^\mathrm{PS} c_k^\dagger 
    c_{2\pi-k}^\dagger\right)\ket{0}.
\end{equation}
Substituting into Eq.~\eqref{bogo} and imposing the condition~\eqref{esc}, one 
obtains $u_k^\mathrm{PS}=\sin\eta_k$ and $v_k^\mathrm{PS}=\cos\eta_k$. Since $\ket{\mathrm{PS}}$ is also 
a Gaussian state, Eq.~\eqref{HH} applies, and we have that
\begin{equation}
    \mathcal{G}_\mathrm{PS}(k) =\mathcal{G}(k,\infty)= -(\cos 2\eta_k +i\sin2\eta_k)=-\frac{G_k+iF_k}{\sqrt{G_k^2+F_k^2}},\label{symevol}
\end{equation}

Recall that in the Ising language, the spin-chain has a $\mathbb{Z}_2$ symmetry generated by $\prod_i \sigma^z_i$. Previously, we considered the $x$-correlator~(\ref{eq:x-corr}) to diagnose spontaneous breaking of this symmetry. As $\sigma^y_i$ does not commute with the same symmetry generator, long-range order in the $y$-correlator 
\begin{equation}
    \rho^y_{n} = \langle \psi | \sigma_i^y \sigma_{i+n}^y | \psi \rangle,
\end{equation}
is an alternative diagnostic for spontaneous symmetry breaking, i.e., it is possible that the system develops long-range ordering of spins in the $y$-basis. The symbol of the above $y$-correlator is related to that of the string correlator by $\mathcal{G}^y(k,\beta) = -e^{ik}\,\mathcal{G}(k,\beta)$. The factor of $e^{ik}$ shifts the winding number by one: a string-correlator symbol with winding number $-1$ corresponds to a $y$-correlator symbol with winding number $0$, which in turn implies that $\rho^y_n$ saturates to a non-zero value at large $n$.

For $\alpha > 1$, the symbols are smooth and non-vanishing on the unit circle, so their winding numbers are well-defined integers. We thus use the winding number, denoted $W[\mathcal{G}]$, to map out the phases of the projection state in parameter space. For the long-range hopping Hamiltonian, this takes the form
\begin{equation}
    \mathcal{G}_{\mathrm{PS}}(k)=-\frac{\mathrm{Li}_\alpha(e^{ik})+\mu}{|\mathrm{Li}_\alpha(e^{ik})+\mu|}.
    \label{lrwn}
\end{equation}
where $\mathrm{Li}_\alpha$ denotes the polylogarithm of order $\alpha$.
Writing $\mathrm{Li}_\alpha(e^{ik}) = A(k) + iB(k)$ with
\begin{equation}
    A(k) = \sum_{n=1}^\infty \frac{\cos(kn)}{n^\alpha}, \qquad B(k) = \sum_{n=1}^\infty \frac{\sin(kn)}{n^\alpha},
\end{equation}
the winding number is obtained by following how the symbol winds around the unit circle as $k$ varies over $[0,2\pi]$. 
The curve intersects the real axis whenever $B(k)=0$. One can verify that $B(k)$ has no zeros in the open intervals $(0,\pi)$ and $(\pi,2\pi)$, therefore the only such intersections occur at $k=0,\pi,2\pi$. At these momenta, the symbol lies at one of the two points $\pm1$ on the unit circle, with the sign fixed by $A(k)+\mu$. For $\mu<-\mathrm{Li}_{\alpha}(1)$, the symbol visits the sequence $+1\to+1\to +1$ at $k=0,\pi,2\pi$, for $\mu>-\mathrm{Li}_{\alpha}(1)$ and $\mu<-\mathrm{Li}_{\alpha}(-1)$ it visits $-1\to +1\to -1$, while for $\mu>-\mathrm{Li}_{\alpha}(-1)$ it visits $-1\to -1\to -1$. 
It remains to determine whether the curve lies in the upper or lower half-plane on each sub-interval. For that, we use the behaviour of the polylog near these points to understand the orientation of the paths. For small $k>0$, one has $B(k)>0$ and therefore $-B(k)<0$. For $k$ near $2\pi$ from below one instead has $B(k)<0$, and therefore $-B(k)>0$.
Combined with the three orderings established above, this determines
the winding numbers. For $\mu<-\mathrm{Li}_\alpha(1)$, the symbol stays
near $+1$: winding
number $0$. For $-\mathrm{Li}_\alpha(1)<\mu<-\mathrm{Li}_\alpha(-1)$,
the symbol moves from $-1$ to $+1$ through the lower half-plane and
returns from $+1$ to $-1$ through the upper half-plane, corresponding
to one clockwise loop around the origin: winding number $+1$ (as an example, see 3rd plot of Fig. \ref{fig:symbol_three_panels_a15} for $\alpha=1.5,\mu = 0$ case). For
$\mu>-\mathrm{Li}_\alpha(-1)$, the symbol stays near $-1$ and again
does not enclose the origin: winding number $0$. Hence a topological phase arises only when $-\mathrm{Li}_\alpha(1) < \mu < -\mathrm{Li}_\alpha(-1)$, in which case the $x$-correlator has winding number $W[\mathcal{G}_\mathrm{PS}^x] = 0$.

For the short-range hopping Hamiltonian, the symbol takes the form 
\begin{equation}
    \mathcal{G}_{\mathrm{PS}}(k)=\frac{t\cos k-\mu-i\,B(k)}{\sqrt{(t\cos k - \mu)^2+B(k)^2}}\,,
\end{equation}
where $B(k) = \mathrm{Im}(\mathrm{Li}_\alpha(e^{ik}))$ is the same function as in the 
long-range case. If $|\mu|>|t|$, the real part $t\cos k - \mu$ has a definite sign for all 
$k$, so the symbol is confined to either the left or right half of the unit circle and 
therefore cannot wind around the origin, giving $W[\mathcal{G}_\mathrm{PS}]=0$, and therefore a trivial phase.

We now deal with the case $|\mu|<|t|$. The curve crosses the real axis only at 
$k = 0, \pi, 2\pi$, where the sign of the real part $t\cos k - \mu$ is determined 
by the sign of $t - \mu$. For $\mu > t$, the real part is negative at $k = 0, 2\pi$ 
and positive at $k = \pi$, so the symbol starts at $-1$, reaches $+1$ at $k = \pi$, 
and returns to $-1$ at $k = 2\pi$. Since $-B(k) < 0$ on $(0,\pi)$ and $-B(k) > 0$ 
on $(\pi, 2\pi)$, the curve travels counter-clockwise through the lower half-plane 
from $-1$ to $+1$ and through the upper half-plane from $+1$ back to $-1$, giving 
$W[\mathcal{G}_\mathrm{PS}] = 1$. For $\mu < t$, the signs are reversed and the symbol traces 
the path $1\to-1\to1$ clockwise, giving $W[\mathcal{G}_\mathrm{PS}] = -1$. In both cases, the 
system is in a topological phase: 
$W[\mathcal{G}_\mathrm{PS}]=1$ implies that the $x$-correlator saturates to a non-zero value, 
while $W[\mathcal{G}_\mathrm{PS}]=-1$ implies the same for the $y$-correlator.

We now turn to the regimes in which the symbol develops a discontinuity. Such discontinuities lead to a logarithmic growth of the entanglement entropy and power-law decay of correlations. They can originate in three distinct ways: from a divergence of the single-particle energy $\epsilon_k$ at some momentum $k$, from a zero of $\epsilon_k$, or from a finite discontinuity of $F_k$ even when $\epsilon_k$ remains finite and nonzero. We discuss these cases in turn. In general, these mechanisms do not always produce a discontinuity in the symbol (unless $\lim_{ k\to k^* }|F_k/G_k|\neq  0$). However, when it happens, the state develops logarithmic entanglement growth and algebraically decaying correlations.

The first mechanism occurs only at $k=0$ for the Hamiltonians considered here, and it occurs whenever $\alpha < 1$, regardless of whether the hopping is short- or long-range. 
In this regime, $\mathrm{Li}_\alpha(e^{ik})$ diverges as $k\to 0$, which dominates the 
symbol and produces the discontinuity
\begin{equation}
    \lim_{k\to 0^\pm}\mathcal{G}_\mathrm{PS}(k)=
    \begin{cases}        -\sin\!\left(\frac{\pi \alpha}{2}\right)\,\mp\, i\,\cos\!\left(\frac{\pi \alpha}{2}\right), & \text{long-range hopping,}\\[6pt]
        \mp\, i, & \text{short-range hopping,}
    \end{cases}
\end{equation}
which holds for all values of $t$ and $\mu$, since the divergence of $\epsilon_k$ 
overwhelms any finite parameter in the Hamiltonian.

The second mechanism is controlled by $\mu$: tuning it so that $\epsilon_k = \sqrt{G_k^2 + F_k^2}$ 
vanishes at some $k$ can produce a discontinuity in the symbol. Since $F_k$ vanishes only at $k = 0$ 
or $k = \pi$, it suffices to tune $\mu$ so that $G_k$ also vanishes at one of these points. 
To make $\epsilon_k = 0$ at $k = \pi$, one requires $\mu = -t$ for $H_{\rm SH}$, and $\mu = -A(\pi) = -\mathrm{Re}[\mathrm{Li}_\alpha(-1)]$ for $H_{\rm LH}$. In both cases this produces the discontinuity
\begin{equation}
    \lim_{k\to \pi^\pm}\mathcal{G}_\mathrm{PS}(k)=\pm i.
\end{equation}
To make $\epsilon_k = 0$ at $k = 0$, one must choose $\alpha>1$ and $\mu = t$ for  $H_{\rm SH}$, and $\mu = -\mathrm{Li}_\alpha(1)$ for $H_{\rm LH}$. This produces the discontinuity
\begin{equation}
    \lim_{k\to 0^\pm}\mathcal{G}_\mathrm{PS}(k)=
    \begin{cases}
        \sin\!\left(\frac{\pi \alpha}{2}\right)\,\pm\, i\,\cos\!\left(\frac{\pi \alpha}{2}\right), & \text{long-range hopping, } 1<\alpha<2,\\[6pt]
        \mp i, & \text{long-range hopping, } \alpha\geq2,\\[6pt]
        \mp\, i, & \text{short-range hopping,}
    \end{cases}
\end{equation}

The third mechanism occurs exclusively for $H_{\rm SH}$ at $\alpha = 1$, where 
$F_k$ develops a finite discontinuity at $k=0$, so the symbol 
inherits a discontinuity without $\epsilon_k$ either vanishing or diverging. This produces,
for all values of $t$ and $\mu$,
\begin{equation}
    \lim_{k\to \pi^\pm}\mathcal{G}_\mathrm{PS}(k)=\frac{t-\mu\mp i\frac{\pi}{2}}{\sqrt{(t-\mu)^2+\left(\frac{\pi}{2}\right)^2}}.
\end{equation}

Once the discontinuities of the symbol have been identified, the effective central charge can be obtained from Eq.~\eqref{genEntropy} by summing the contribution of each jump. More explicitly, for every discontinuity point $k^*$, one evaluates the imaginary part of the symbol on the two sides of the jump,
\[
g_{k^*}^\pm=\lim_{k\to k^{*\pm}} \mathrm{Im}\,\mathcal{G}_\mathrm{PS}(k) ,
\]
and inserts the corresponding jump data into Eq.~\eqref{genEntropy}. The total effective central charge is then obtained by adding the contributions from all discontinuity points $k^*$. The power-law decay of correlations can then be obtained 
via Eq.~\eqref{factorizationdisc}, by analyzing the regularized continuous symbol
\begin{equation}
\tilde{\mathcal{G}}_\mathrm{PS}(k)=\mathcal{G}_\mathrm{PS}(k)\prod_{\sigma}e^{-i t_\sigma(k-k_\sigma-\pi\,\mathrm{sgn}(k-k_\sigma))},
\end{equation}
where $t_\sigma = -\arg(\lim_{k\to k_\sigma^+} \mathcal{G}_\mathrm{PS}(k))/\pi + n$, $k_\sigma$ are the discontinuity points, and $n$ is 
an integer chosen such that the winding number of $\tilde{\mathcal{G}}_\mathrm{PS}$ vanishes, 
which can be determined by a similar analysis as the one at the beginning of this appendix.

\section{Instantaneous or asymptotic deformation of the symbol
}\label{app:symbol}
In this appendix, we explain how to identify whether the transition between a trivial and a topological phase or to a critical-like behavior occurs at $\beta=0$ or at $\beta=\infty$.
The general evolved state is given by: 
\begin{equation}
\label{77}
| \psi(\beta)\rangle=c^{\dagger}_0\bigotimes_{k} \vert \psi_k(\beta)\rangle=
c^{\dagger}_0\bigotimes_{k}
\left(\frac{u_k(\beta)+v_k(\beta)c^\dagger_{k} c^\dagger_{2\pi-k}}{\sqrt{| u_k(\beta)\vert^2+\vert v_k(\beta)\vert^2}}\right)| 0\rangle 
\,,
\end{equation}
where
\begin{equation}
    \begin{split}
        u_k(\beta)=&\,\left(u_k(0)-\frac{\tanh (\beta  \epsilon_k)  
   (G_k u_k(0)-F_k v_k(0))}{\epsilon_k}\right),\\
   v_k(\beta)=\, & \left(v_k(0) +\frac{\tanh (\beta  \epsilon_k)
   (G_k v_k(0)+F_k u_k(0))}{\epsilon_k}\right),
    \end{split}
    \label{apevo}
\end{equation}
where $u_k(0)$ is continuous in $k\in [0,2\pi]$ and $v_k(0)$ can be discontinuous in $k=0$ ($h=1$). Also, we took advantage of the rescaling redundancy of $u_k(\beta)$ and $v_k(\beta)$ to rewrite \eqref{evolutionformula} in this simpler form \eqref{apevo}. $G_k$ and $F_k$ are defined in $k\in (0,2\pi)$.

We first analyze an immediate flow to the topological phase for any $\beta > 0$. It occurs if the phase of the state at exactly $\beta=0$ must differ from the phase at arbitrarily small but positive $\beta$. Since the topological features are encoded in the symbol, this can only happen if the limit $\beta\to0^+$ is not uniform in momentum. Equivalently, there must exist a momentum $k^*$ such that the limits $k\to k^*$ and $\beta\to0^+$ do not commute:
\begin{equation}
    \lim_{k\to k^*} \lim_{\beta\to 0^+} \mathcal{G}(k,\beta) \neq  \lim_{\beta\to 0^+}\lim_{k\to k^*}\mathcal{G}(k,\beta) 
    \label{trans_cond}
\end{equation}
for some $k^*$. If instead these limits commute for every $k^*$, then the symbol changes smoothly as soon as $\beta>0$, and no instantaneous change can take place.

Consider a fixed momentum mode $k$. Taking the limit $\beta\to 0^+$ in Eq.~\eqref{apevo}, the evolved coefficients reduce continuously to their initial values,
$ u_k(0^+)=u_k(0), v_k(0^+)=v_k(0),
$
provided the single-particle energy $\epsilon_k$ remains finite. Therefore, an immediate transition at $\beta=0$ can occur only if $\epsilon_k$ becomes singular at some momentum $k^*$.

As an example, consider the evolution with long-range hopping, $\alpha = 1$,$\mu = 0$, and an initial 
state in the trivial phase, $h = 1.5, \gamma=1$. We focus on the symbol of the string correlator for the projection state; once its winding number is known, the winding numbers of the other correlator symbols follow by additivity. For $k\in(0,2\pi)$,
\begin{equation}
    \mathcal{G}_\mathrm{PS}(k)=-\frac{\mathrm{Li}_1(e^{ik})}{|\mathrm{Li}_1(e^{ik})|}
    =\frac{\log|2\sin(k/2)|-i\tfrac{\pi-k}{2}}
    {\left|\log|2\sin(k/2)|-i\tfrac{\pi-k}{2}\right|}.
    \label{sa1}
\end{equation}
As $k \to 0^+$ or $k \to 2\pi^-$, the real part $\log|2\sin(k/2)|$ diverges to $-\infty$ while the imaginary part remains finite. After normalization, the symbol therefore approaches $\mathcal{G}_\mathrm{PS} \to -1$ at both endpoints. The symbol thus extends continuously across $k \in [0, 2\pi]$, even though $\mathrm{Li}_1(e^{ik})$ itself has a logarithmic singularity at $k = 0$. Near $k=0^+$, the imaginary part $-(\pi-k)/2$ is negative, so the curve approaches $-1$ from below the real axis. Near $k=2\pi^-$, the imaginary part is positive, so the curve approaches $-1$ from above. At $k=\pi$, the imaginary part vanishes and the real part is $\log 2>0$, so $\mathcal{G}_\mathrm{PS}(\pi)=+1$. As $k$ runs from $0^+$ to $2\pi^-$, the curve starts near $-1$ in the lower half-plane, reaches $+1$ at $k=\pi$, and returns to $-1$ through the upper half-plane. In this way, it makes one full counterclockwise turn around the origin:
\begin{equation}
W\left[\mathcal{G}_\mathrm{PS}\right]=1.
\end{equation}
Therefore, the $x$-correlator has winding number
\begin{equation}
W[\mathcal{G}_\mathrm{PS}^x]=0,
\end{equation}
confirming that the projection state is in the topological phase.

At the same time, the quantity
$\epsilon_k = |\mathrm{Li}_1(e^{ik})|$
diverges at $k=0$. This suggests that the change from the trivial to the topological phase takes place already at $\beta=0^+$, rather than only at large $\beta$. Indeed, $F_k$ and $G_k$ are continuous for all $k\in(0,2\pi)$, so for every momentum strictly inside the interval the limits $\beta\to0^+$ and evaluation at fixed $k$ commute, giving $u_k(0^+)=u_k(0)$ and $v_k(0^+)=v_k(0)$. The only possible obstruction is therefore at the edges, $k\to0^+$ and $k\to2\pi^-$, where $\epsilon_k$ diverges and $\tanh(\beta\epsilon_k)$ can remain finite even when $\beta\to0^+$. The transition at $\beta=0$ is thus entirely driven by modes in an arbitrarily small neighborhood of $k=0$.

To make this explicit, let us evaluate Eq.~\eqref{apevo} near the singular momentum $k^*=0$, starting from a trivial initial condition,
$u_{k^*}(0)=1, v_{k^*}(0)=0$.
Then
\begin{equation}
    u_{k\approx k^*}(\beta)
    =
    1-\tanh(\beta\epsilon_{k\approx k^*})\,
    \frac{G_{k\approx k^*}}{\epsilon_{k\approx k^*}},
    \qquad
    v_{k\approx k^*}(\beta)
    =
    \tanh(\beta\epsilon_{k\approx k^*})\,
    \frac{F_{k\approx k^*}}{\epsilon_{k\approx k^*}}.
    \label{eq:evol_trivial}
\end{equation}
Because $\epsilon_{k\approx k^*}$ diverges, we have $\tanh(\beta\epsilon_k)\to 1$ for any fixed $\beta>0$. The behavior of the mode is then controlled by the ratios
\begin{equation}
    \tilde G_{0^+}=\lim_{k\to0^+}\frac{G_k}{\epsilon_k},
    \qquad
    \tilde F_{0^+}=\lim_{k\to0^+}\frac{F_k}{\epsilon_k}.
\end{equation}
In the present case,
\begin{equation}
G_k=-\log|2\sin(k/2)|,
\qquad
F_k=\frac{\pi-k}{2},
\end{equation}
so $G_k$ diverges logarithmically while $F_k$ remains bounded. Therefore
\begin{equation}
\tilde G_{0^+}=1,
\qquad
\tilde F_{0^+}=0.
\end{equation}
Substituting into Eq.~\eqref{eq:evol_trivial}, one finds formally
\begin{equation}
    \lim_{k\to0^+}u_k(\beta)=0,
    \qquad
    \lim_{k\to0^+}v_k(\beta)=0,
    \qquad (\beta>0).
\end{equation}
This apparent vanishing simply means that the leading terms cancel, so one must keep the first subleading corrections. Expanding more carefully,
\begin{equation}
\frac{G_k}{\epsilon_k}\sim 1-\frac{F_k^2}{2G_k^2}+\cdots,
\qquad
\frac{F_k}{\epsilon_k}\sim \frac{F_k}{G_k}\sim \frac{\pi/2}{-\log k},
\end{equation}
which gives
\begin{equation}
    u_k(\beta)\sim \frac{\pi^2/8}{(\log k)^2},
    \qquad
    v_k(\beta)\sim -\frac{\pi/2}{\log k},
    \qquad (k\to0^+,\ \beta>0).
\end{equation}
Hence $|u_k|\ll |v_k|$, so the singular mode is dominated by the paired component for any positive $\beta$, however small.

Recalling that the symbol is $\mathcal{G}(k,\beta) = n(k,\beta) + g(k,\beta)$, with $n$ and $g$ given by Eq.~\eqref{eq:ng}, we obtain
\begin{equation}
    \lim_{\beta\to0^+}\lim_{k\to0^+}\mathcal{G}(k,\beta)
    =
    -1
    \;\neq\;
    \lim_{k\to0^+}\lim_{\beta\to0^+}
    \mathcal{G}(k,\beta)
    =
    1.
    \label{eq:limit_ng}
\end{equation}
The two orders of limits therefore, give different answers. If one takes $k\to0^+$ first, the singular mode is immediately driven to the projection-state value, corresponding to $\mathcal{G}_\mathrm{PS}(0^+)=-1$. If instead one takes $\beta\to0^+$ first, one recovers the initial value of the trivial phase. This shows explicitly that the phase transition is already present at $\beta=0^+$.

\begin{figure}[t!]
    \centering
    \includegraphics[width=\textwidth]{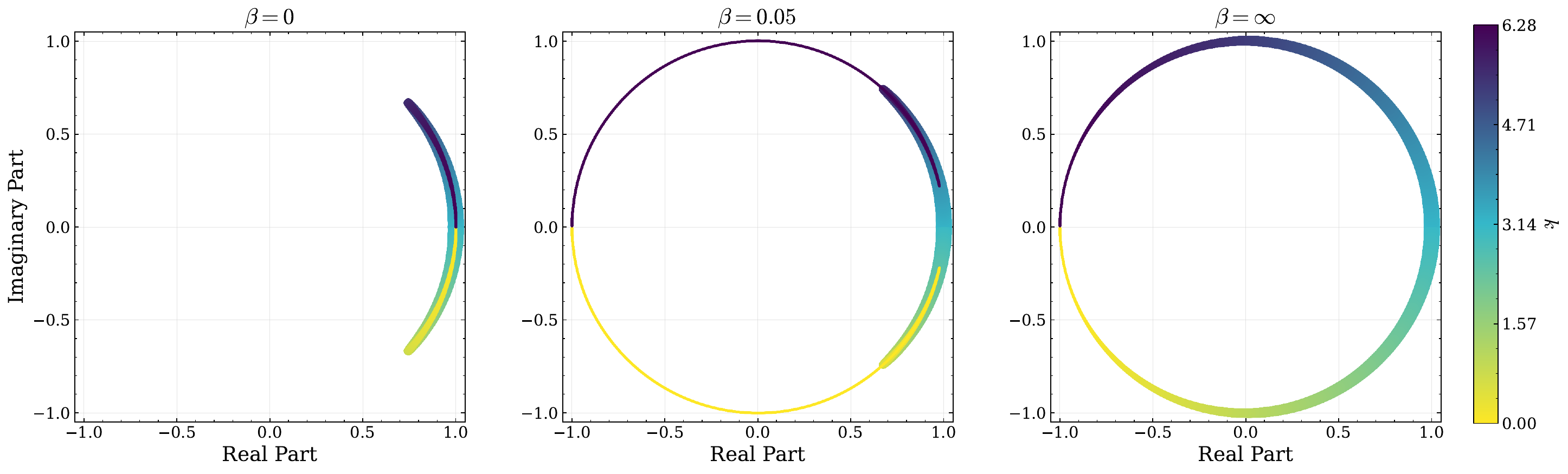}
    \caption{\textbf{Evolution of the correlator symbol across $\beta$ for 
    $\alpha = 1$.} The symbol $\mathcal{G}(k,\beta) = (n(k,\beta)+g(k,\beta))$ is 
    shown in the complex plane for the long-range hopping Hamiltonian with 
    $\alpha = 1,\mu = 0$ and initial state $h = 1.5, \gamma = 1$ (trivial phase). 
    Left: at $\beta = 0$ the symbol has winding number zero around the origin. 
    Middle: at $\beta = 0.05$ the symbol has already wound around the origin, 
    with the modes near $k = 0$ pulled to the projection-state value while 
    modes away from the singularity still reflect the initial state. Right: 
    at $\beta \to \infty$ the symbol coincides with the projection state and 
    traces the full unit circle with winding number $-1$. The color encodes $k \in (0, 2\pi)$, and point size is scaled with $|k - \pi|$ 
    so that overlapping points remain visible.}
    \label{fig:symbol_three_panels}
\end{figure}

This behavior is illustrated in Fig.~\ref{fig:symbol_three_panels}, where we show the symbol at $\beta=0$, at a small finite value $\beta=0.05$, and in the projection limit $\beta\to\infty$. Only the modes in a small neighborhood of $k=0^+$ and $k=2\pi^-$ are modified at small $\beta$, but this localized rearrangement is already sufficient to change the winding number from $0$ to $-1$.

We now contrast this with the case $\alpha >1$ discussed in the main text. Here $\epsilon_k$ is 
bounded for every mode, so the mechanism responsible for the $\beta = 0$ transition is absent. Moreover, $F_k$ and $G_k$ are continuous on the entire 
interval $k \in [0, 2\pi]$, which ensures that
\begin{equation}
    \lim_{k\to k^*} \lim_{\beta\to 0^+} \mathcal{G}(k,\beta) = \lim_{\beta\to 0^+}\lim_{k\to k^*}\mathcal{G}(k,\beta) 
    \label{trans_cond_a12}
\end{equation}
for all $k^* \in [0, 2\pi]$, confirming the absence of a transition at $\beta = 0$. 

Nevertheless, the projection state is topological, as shown in 
Appendix~\ref{app:Hev}. A transition between the trivial initial state at 
$h > 1$ and the topological projection state must therefore occur at some 
finite value of $\beta$ along the evolution, or be pushed to $\beta \to \infty$. 
The evolution in Eq.~\eqref{apevo} is continuous in both $\beta$ and $k$ on 
$[0, 2\pi]$, and no finite-$\beta$ deformation of a continuous symbol with unit modulus can change its winding number. The transition is therefore 
necessarily realized only in the limit $\beta \to \infty$. We must therefore have that
\begin{equation}
    \lim_{k\to k^*} \lim_{\beta\to \infty} 
    \mathcal{G}(k,\beta) \neq  \lim_{\beta\to \infty}\lim_{k\to k^*}\mathcal{G}(k,\beta)
    \label{trans_condinf}
\end{equation}
The demonstration will be analogous to the $\beta=0$ transition. The non-commutativity of limits arises at the isolated mode where $F_k$ 
vanishes, namely $k^* = 0$. At exactly $k^* = 0$, the condition $F_{k^*} = 0$ 
means that $v_{k^*}(\beta) = 0$ for every $\beta$, while $u_{k^*}(\beta) = 
1 - \tanh(\beta\epsilon_{k^*}) G_{k^*}/\epsilon_{k^*}=1 - \tanh(\beta\epsilon_{k^*})$ remains strictly 
positive. The normalized mode is therefore pinned at $(1, 0)$ throughout 
the evolution, and in particular 
$$\lim_{\beta\to\infty}(u_{k^*}(\beta), v_{k^*}(\beta))/
\|u_{k^*}^2(\beta)+v_{k^*}^2(\beta)\| = (1, 0).$$

When we exchange the order of the limits, near $k=0$ both $u_k(\infty)$ 
and $v_k(\infty)$ vanish but at different rates: 
$u_k(\infty) \sim F_k^2/(2 G_k^2)$ while $v_k(\infty) \sim F_k/G_k$, with 
$|u_k| \ll |v_k|$. The symbol satisfies
\begin{equation}
    \lim_{\beta\to\infty}\lim_{k\to 0} 
    \mathcal{G}(k,\beta) 
    =  1 
    \;\neq\; 
    \lim_{k\to 0}\lim_{\beta\to\infty} 
    \mathcal{G}(k,\beta)  
    = -1.
    \label{eq:limit_ng_a15}
\end{equation}
\begin{figure}[t!]
    \centering
    \includegraphics[width=\textwidth]{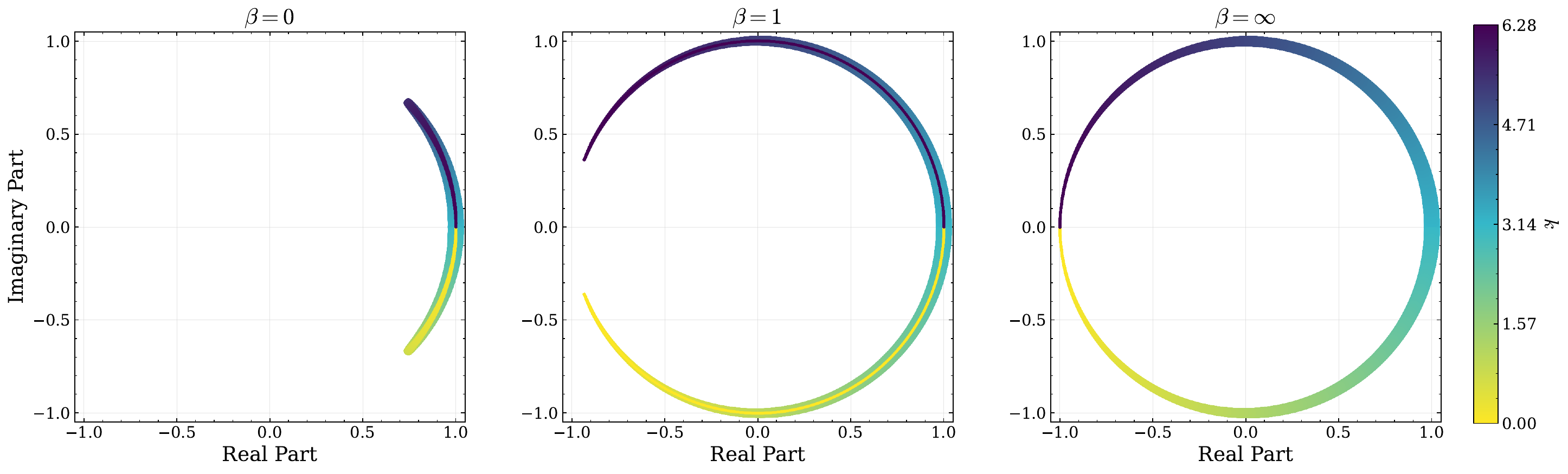}
    \caption{\textbf{Evolution of the correlator symbol across $\beta$ for 
    $\alpha = 1.5$.} The symbol $\mathcal{G}(k,\beta) = (n(k,\beta)+g(k,\beta))$ 
    is shown in the complex plane for the long-range hopping Hamiltonian with 
    $\alpha = 1.5$ and initial state $h = 1.5, \gamma = 1$ (trivial phase). 
    Left: at $\beta = 0$ the symbol has winding number zero. Middle: at 
    $\beta = 1$, the symbol has deformed continuously and still carries 
    winding number zero, with the mode at $k = 0$ pinned near the initial 
    value while the surrounding modes have rotated towards the projection 
    state. Right: at $\beta \to \infty$ the symbol coincides with the 
    projection state and wraps the unit circle with winding number $-1$, 
    with the projection-state values at $k = 0^\pm$ approached tangentially 
    from opposite half-planes. The color encodes the momentum $k \in (0, 2\pi)$, 
    and point size is scaled with $|k - \pi|$ so that the tangential approach 
    at $k = 0, 2\pi$ remains visible.}
    \label{fig:symbol_three_panels_a15}
\end{figure}
This behavior is illustrated in Fig.~\ref{fig:symbol_three_panels_a15}. 
Unlike the $\alpha = 1$ case, the middle panel at $\beta = 1$ shows a symbol 
that has deformed smoothly across the whole interval 
without changing its winding number, consistent with the absence of a 
finite-$\beta$ transition. One can think of the evolution as continuously 
carrying every mode from the initial state to the projection state. The 
transition is pushed to $\beta \to \infty$ because modes arbitrarily close 
to $k = 0$ requires an arbitrarily large $\beta$ to effectively move from the initial to the final value.

We now discuss a different type of asymptotic flow as $\beta\to\infty$, in which the system does not approach a gapped topological phase, but instead flows to a state characterized by logarithmic entanglement growth and power-law correlations. This behavior can arise when the evolution Hamiltonian is gapless, i.e., when $\epsilon_k=0$ for some momentum $k$. The parameter regimes where this occurs are detailed in Appendix~\ref{app:Hev}, where we also show that the corresponding projection state exhibits logarithmic entanglement and algebraic correlations.

As a concrete example, consider an initial trivial state with $h=1.5$, $\gamma=1$, and a deformation with short-range hopping and chemical potential $\mu=-t$. As discussed in Appendix~\ref{app:Hev}, this choice produces a discontinuity in the projection-state symbol at $k=\pi$, where $F_\pi=G_\pi=0$ and hence $\epsilon_\pi=0$. We now verify that condition~\eqref{trans_condinf} is satisfied at $k^*=\pi$.

Taking first the limit $\beta\to\infty$, Eq.~\eqref{apevo} yields
\begin{equation}
\begin{split}
u_k(\infty)
&=
u_k(0)-
\frac{G_k}{\epsilon_k} u_k(0)+\frac{F_k}{\epsilon_k}v_k(0),\\
v_k(\infty)
&=
v_k(0)+\frac{G_k}{\epsilon_k} v_k(0)+\frac{F_k}{\epsilon_k} u_k(0).
\end{split}
\end{equation}
Near $k=\pi$, using $G_k=-t(1+\cos k)$ and $F_k=\mathrm{Im}[\mathrm{Li}_\alpha(e^{ik})]$, one finds
\[
\frac{G_k}{\epsilon_k}\to 0,
\qquad
\frac{F_k}{\epsilon_k}\to \mp 1
\quad\text{as } k\to\pi^\pm.
\]
With initial conditions $u_\pi(0)=1$, $v_\pi(0)=0$, this gives
\begin{align}
\lim_{k\to\pi^\pm}\lim_{\beta\to\infty}u_k(\beta) &= 1,\\
\lim_{k\to\pi^\pm}\lim_{\beta\to\infty}v_k(\beta) &= \mp 1,
\end{align}
so that $n+g=\pm i$. In contrast, exactly at $k=\pi$ one has $F_\pi=G_\pi=0$, and Eq.~\eqref{apevo} fixes the mode to its initial value for all $\beta$: $u_\pi(\beta)=1$, $v_\pi(\beta)=0$, giving $n+g=1$ even after taking $\beta\to\infty$. The two limits therefore, do not commute:
\begin{equation}
\lim_{k\to\pi^\pm}\lim_{\beta\to\infty}\mathcal{G}(k,\beta)
\;=\;\pm i
\;\neq\;
\lim_{\beta\to\infty}\lim_{k\to\pi^\pm}\mathcal{G}(k,\beta)
\;=\;1,
\end{equation}
confirming that condition~\eqref{trans_condinf} is satisfied at $k^*=\pi$.

The discontinuity in the symbol is thus generated only in the strict $\beta\to\infty$ limit. Modes arbitrarily close to $k=\pi$ have $\epsilon_k\to0$ and therefore require arbitrarily large $\beta$ to evolve away from their initial values, approaching the projection state from opposite sides of the unit circle. 

\begin{figure}[t!]
    \centering
    \includegraphics[width=\textwidth]{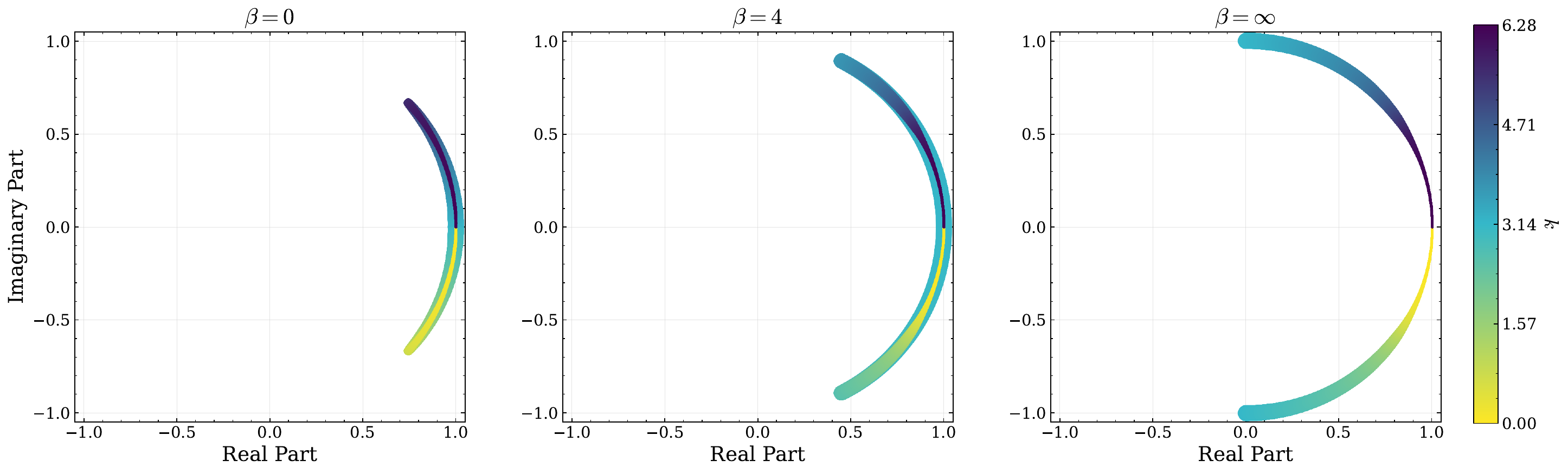}
    \caption{\textbf{Evolution of the correlator symbol across $\beta$ for 
    the gapless evolution with $\mu = -t$.} The symbol 
    $\mathcal{G}(k,\beta) = (n(k,\beta)+g(k,\beta))$ is shown in the complex 
    plane for the short-range hopping Hamiltonian with $\mu = -t$,$\alpha = 1.5$, and 
    initial state $h = 1.5, \gamma = 1$ (trivial phase). Left: at 
    $\beta = 0$, the symbol lies on a small arc around the initial value 
    $\mathcal{G}(0,0) = 1$. Middle: at $\beta = 4$, the arc has opened up and 
    modes away from $k = \pi$ have rotated significantly towards the 
    projection state, while a neighborhood of $k = \pi$ (mid-range colors) 
    is still pinned near $1$. Right: at $\beta \to \infty$ the symbol 
    traces the unit circle with a discontinuity at $k = \pi$, where the 
    left and right limits give $+i$ and $-i$ respectively. The color encodes 
    $k \in (0, 2\pi)$, and point size is scaled with $|k - \pi|$ so that 
    overlapping points remain visible.}
    \label{fig:symbol_three_panels_mu}
\end{figure}

This mechanism is illustrated in Fig.~\ref{fig:symbol_three_panels_mu}, where we show the symbol at $\beta=0$, at an intermediate value $\beta=4$, and in the $\beta\to\infty$ limit. While the projection-state symbol becomes discontinuous, jumping between $-i$ and $+i$ at $k=\pi$, no such discontinuity exists at any finite $\beta$. Instead, the evolution progressively deforms the symbol away from the initial state, with a shrinking neighborhood around $k=\pi$ remaining pinned to its initial value. The jump emerges only asymptotically, reflecting the vanishing of $\epsilon_k$ at the critical value of the momentum.

Our final example also evolves to a critical-like state, but the transition
now occurs at $\beta = 0$. This happens when $F_k$ diverges at some
$k = k^*$ while the ratio $G_k/F_k$ remains bounded as $k \to k^*$, which
in particular forces $\epsilon_k$ to diverge at $k^*$. The precise
parameter choices realizing this scenario are described in
Appendix~\ref{app:Hev}, where we also show that the corresponding
projection state exhibits logarithmic entanglement growth and power-law
correlations.
 
We now evaluate \eqref{apevo} near the singular point $k^* = 0$ with the
trivial-phase initial condition $u_{k^*}(0) = 1$, $v_{k^*}(0) = 0$:
\begin{equation}
    u_{k\approx k^*}(\beta) = 1 - \tanh(\beta\epsilon_{k\approx k^*})\,
    \frac{G_{k\approx k^*}}{\epsilon_{k\approx k^*}},
    \qquad
    v_{k\approx k^*}(\beta) = \tanh(\beta\epsilon_{k\approx k^*})\,
    \frac{F_{k\approx k^*}}{\epsilon_{k\approx k^*}}.
    \label{eq:evol_trivial_a05}
\end{equation}
Since $\epsilon_{k\approx k^*}$ diverges, $\tanh(\beta\epsilon_k) \to 1$
for any $\beta > 0$, and the limiting values depend on the ratios
$G_k/\epsilon_k$ and $F_k/\epsilon_k$. Taking $\alpha<1$ with $H_{SH}$ as an example, these give
$F_k/\epsilon_k \to 1$ and $G_k/\epsilon_k \to 0$ at $k \to 0^+$, so from
\eqref{eq:evol_trivial_a05}
\begin{equation}
    \lim_{k\to 0^+}u_k(\beta) = 1, \qquad
    \lim_{k\to 0^+}v_k(\beta) = 1 \qquad (\beta > 0).
\end{equation}
We therefore have 
\begin{equation}
    \lim_{\beta\to 0^+}\lim_{k\to 0^+}
    \mathcal{G}(k,\beta) = -i
    \;\neq\;
    \lim_{k\to 0^+}\lim_{\beta\to 0^+}
    \mathcal{G}(k,\beta) = 1,
\end{equation}
so \eqref{trans_cond} is satisfied at $k^* = 0$ and the transition is
generated instantaneously at $\beta = 0^+$ by the modes in an
arbitrarily small neighborhood of the singularity.

\begin{figure}[t!]
    \centering
    \includegraphics[width=\textwidth]{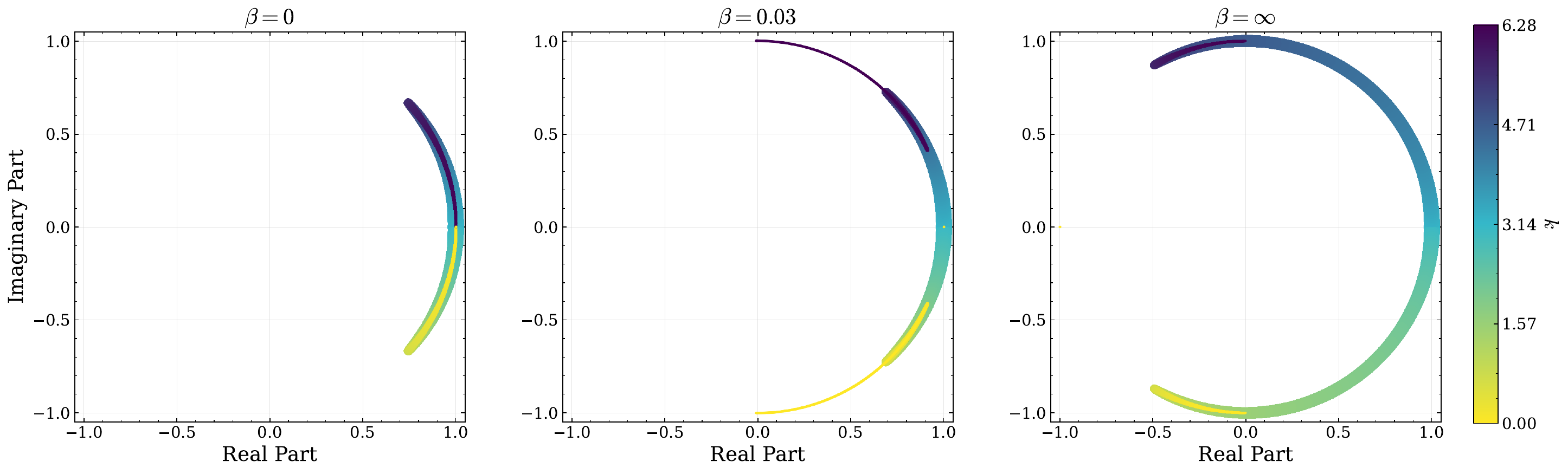}
    \caption{\textbf{Evolution of the correlator symbol across $\beta$
    for the critical-like transition at $\alpha = 0.5$.} The symbol
    $\mathcal{G}_\mathrm{PS}(k) = (n(k,\beta)+g(k,\beta))$ is shown in the complex
    plane for the long-range hopping Hamiltonian with $\alpha = 0.5$,
    $t = -1$, $\mu = 0$, and trivial initial state $h = 1.5$,
    $\gamma = 1$. Left: at $\beta = 0$ the symbol reflects the trivial
    initial state, an arc near $\mathcal{G}_\mathrm{PS} = 1$ with winding number
    zero. Middle: at $\beta = 0.03$ the modes near the singularity
    $k = 0$ (yellow and dark purple ends of the colormap, corresponding
    to $k \to 0^+$ and $k \to 2\pi^-$) have already been pulled to the
    projection-state values $\pm i$, while the remaining modes still
    trace an arc close to the initial state. Right: at $\beta \to \infty$
    the symbol coincides with the projection state, which traces most
    of the unit circle, but develops a discontinuity at $k = 0$, with
    the two one-sided limits $\pm i$ approached tangentially from
    opposite half-planes. The color encodes $k \in (0, 2\pi)$.}
    \label{fig:symbol_three_panels_a05}
\end{figure}

This behavior is illustrated in Fig.~\ref{fig:symbol_three_panels_a05},
where we plot the symbol at $\beta = 0$, at a small value $\beta = 0.03$,
and at $\beta \to \infty$ (projection state). As in the $\alpha = 1$ case,
only the modes in an arbitrarily small neighborhood of $k = 0$ are
affected at small $\beta$, yet this localized change is already
sufficient to drive the symbol to the projection-state values $\pm i$
at $k = 0^\pm$ and generate the discontinuity responsible for the
critical-like behavior. Unlike our first example, the projection
state is not a smooth closed curve but develops a discontinuity at
$k = 0$: the symbol approaches $+i$ as $k \to 2\pi^-$ and $-i$ as
$k \to 0^+$.
\end{document}